\documentclass[11pt]{article}

\usepackage[final]{acl}
\usepackage{placeins}
\usepackage{hyperref}
\usepackage{url}
\usepackage{graphicx}
 \usepackage{booktabs}
\usepackage{tabularx}
\usepackage{paralist}
\usepackage{multirow}
\usepackage{comment}
\usepackage{adjustbox}
\usepackage{times}
\usepackage{latexsym}
\usepackage[table]{xcolor}   % for row colors
\usepackage{enumitem}
\usepackage{listings}
\usepackage{float} 
\usepackage[T1]{fontenc}
\usepackage[scaled]{beramono} 

\usepackage{xurl}

\usepackage[utf8]{inputenc}

\usepackage{microtype}

\usepackage{inconsolata}

\title{Beyond Code Pairs: Dialogue-Based Data Generation for LLM Code Translation}

\makeatletter
\def\@titleheader{Beyond Code Pairs}
\def\@authorrunning{Caiwen Ding (UMN) and Chunhua ``Leo'' Liao (LLNL)}
\makeatother

\author{
Le Chen$^{1}$\thanks{Equal contributions.},
Nuo Xu$^{2}$\footnotemark[1],
Winson Chen$^{2}$,
Bin Lei$^{2}$,
Pei-Hung Lin$^{4}$,\\
\textbf{Dunzhi Zhou}$^{2}$,
\textbf{Rajeev Thakur}$^{1}$,
\textbf{Caiwen Ding}$^{2}$,
\textbf{Ali Jannesari}$^{3}$,
\textbf{Chunhua Liao}$^{4}$\\
$^{1}$Argonne National Laboratory,
$^{2}$University of Minnesota,\\
$^{3}$Iowa State University,
$^{4}$Lawrence Livermore National Laboratory\\
\texttt{lechen@anl.gov, liao6@llnl.gov}
}

\begin{document}
\maketitle

\begin{abstract}
Large language models (LLMs) have shown remarkable capabilities in code translation, yet their performance deteriorates in low-resource programming domains such as Fortran and emerging frameworks like CUDA, where high-quality parallel data are scarce. We present an automated dataset generation pipeline featuring a dual-LLM Questioner–Solver design that incorporates external knowledge from compilers and runtime feedback. Beyond traditional source–target code pair datasets, our approach additionally generates (1) verified translations with unit tests for assessing functional consistency and (2) multi-turn dialogues that capture the reasoning process behind translation refinement. 
Applied to Fortran→C++ and C++→CUDA, the pipeline yields 3.64k and 3.93k dialogues, respectively. 
Fine-tuning on this data yields dramatic improvements in functional correctness, boosting unit test success rates by over 56\% on the challenging C++-to-CUDA task. We show that the generated data enables a 7B open-weight model to significantly outperform larger proprietary systems on key metrics like compilation success.\let\thefootnote\relax\footnotetext{\raggedright Code and data: \url{https://github.com/HPC-Fortran2CPP/beyond-code-pairs-translation}}
\end{abstract}
% 1. motivation
% 2. Challenges of LLM approaches
% 3. Limitations of existing datasets
% 4. Our approach
% 5. contr.
\section{Introduction}
Automated code translation has long been a goal in programming languages and systems research, enabling developers to migrate software across languages, frameworks, and hardware platforms~\cite{ahmad2021unified, feng2020codebert, lachaux2020unsupervised}. In high-performance computing (HPC) and scientific computing, this problem is particularly pressing since legacy codes need to interoperate with modern ecosystems while new frameworks such as CUDA and OpenMP continue to emerge to exploit specialized hardware~\cite{czarnul2020survey, dhruv2025leveraging}. Large language models (LLMs) have recently demonstrated impressive capabilities in code understanding and generation, motivating their application to code translation. Yet their effectiveness is uneven, with good results in popular languages such as Python and Java, but substantially weaker performance in low-resource domains and specialized frameworks where training data is scarce and structural complexity is high. For example, LLMs often produce translations that compile but fail unit tests or generate code that is syntactically invalid without guidance from external tools~\cite{chen2024landscape}.

\begin{figure}[!t]
\centering
\includegraphics[width=0.4\textwidth]{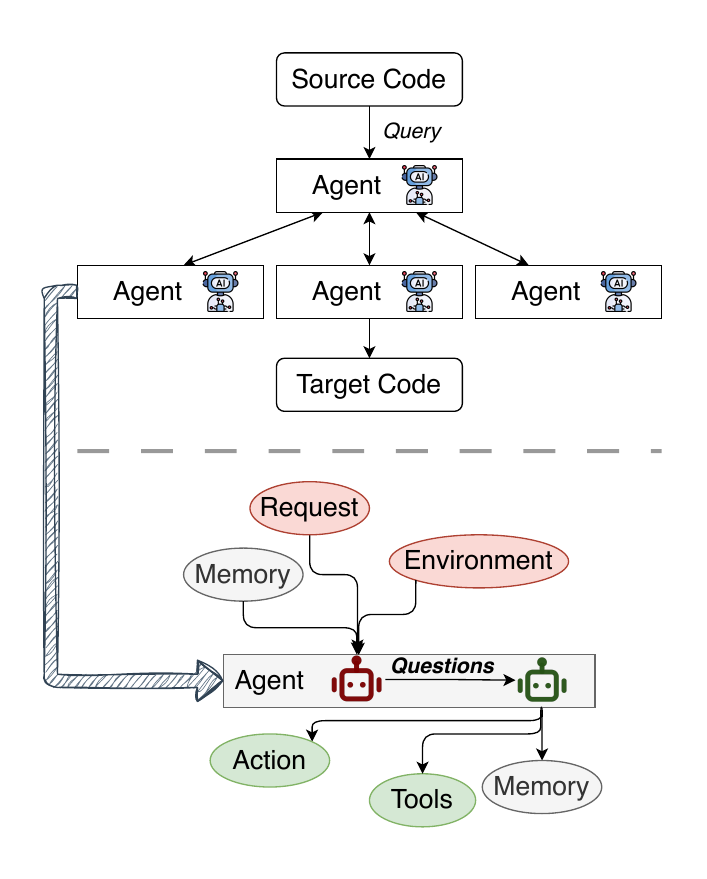} 
\caption{Replacing the single-LLM core inside each agent with a dual-LLM Questioner–Solver core enables explicit reasoning separation and tool-grounded refinement.}
\label{fig:agent}
\end{figure}

% Existing approaches face two key challenges. First, the scarcity of parallel datasets in low-resource languages and frameworks limits the ability of LLMs to learn accurate mappings. Generating or verifying data requires a significant amount of labor. Second, traditional datasets focus on static source–target pairs, thus failing to fully exploit LLMs' capability in capturing the reasoning and error-correction workflows required in practice.  Addressing these gaps requires a framework that can both generate scalable, high-quality training data and embed richer reasoning signals into the learning process.

Agent-based approaches have been widely adopted to address the limitations of LLMs in code translation. Most existing systems follow a source-to-target workflow~\citep{dearing2024lassi, bitan2025unipar,chen2025pcebench}, where multiple agents iteratively perform translation, verification, and refinement until a syntactically valid target program is produced (Figure \ref{fig:agent}, up). The resulting outputs are typically source–target code pairs, which are well-suited for traditional supervised learning pipelines~\cite{chen2021evaluating}. Although these approaches can produce correct translations, they discard valuable intermediate information that could further enhance the performance of LLMs. This information, such as compiler feedback, runtime diagnostics, and interactions with external tools or scripts, has been shown to be crucial for enabling deeper reasoning about program semantics and functional behavior~\citep{ding2024semcoder}.

% However, these approaches largely treat LLMs as black boxes. Although they can yield correct translations, they fail to capture or exploit the reasoning steps that LLMs naturally generate when diagnosing compiler errors, interacting with tools, or applying fixes. Consequently, current pipelines overlook the opportunity to transform these intermediate traces into structured supervision that could enhance both interpretability and downstream performance.

% https://drive.google.com/drive/folders/1q5uok7TbeSeTgwuv5BEesIcPkmY2SNQd?usp=sharing

To address this gap, we introduce a dual-LLM Questioner–Solver design within each LLM agent, which automatically collects and generates reasoning process data during the agent-based translation workflow (Figure \ref{fig:agent}, bottom). Instead of producing only final code pairs, our framework automatically generates question-solution data covering intermediate queries, responses, and tool interactions that occur during translation. The Questioner analyzes the current state, formulates targeted questions, and incorporates external signals such as compiler errors or runtime outputs, while the Solver executes translations, generates unit tests, and proposes fixes. This interaction yields multi-turn dialogues that not only capture verified source–target programs but also encode the decision-making and error-correction steps behind them. By embedding this structured reasoning into the dataset, our approach enables fine-tuned LLMs to learn both the end results of translation and the process by which correctness is achieved.

% To address these challenges, we introduce a framework that can both generate scalable, high-quality data and embed richer reasoning signals into the learning process. At the core of our approach is a dual-LLM Questioner–Solver module, where one agent analyzes intermediate states and formulates queries while the other executes translations, generates unit tests, and repairs errors. The pipeline integrates compilers, runtime environments, and custom scripts, embedding their feedback into structured reasoning dialogues from the interactions between Questioners and Solvers. Rather than discarding intermediate failures, the process transforms them into training signals, enabling fine-tuned LLMs to learn not just end-to-end mappings but also the iterative problem-solving strategies behind translation.

Our contributions are as follows:

\begin{itemize}[nosep]
    \item An automated pipeline for code translation: we introduce a multi-turn dialogue generation framework driven by a dual-LLM Questioner–Solver module that integrates external knowledge into structured reasoning traces.
    \item Two new translation benchmarks: we instantiate this pipeline on Fortran→C++ and C++→CUDA, producing large-scale datasets of 3.64k and 3.93k dialogues with verified translations and unit tests. These are further decomposed into over \textbf{14.1k} (Fortran→C++) and \textbf{12.7k} (C++→CUDA) Question-Solution pairs and are released alongside three disjoint test suites for rigorous evaluation.
    % \item Demonstrated improvements in low-resource translation: Fine-tuning open-weight LLMs on our datasets yields up to 3.31× improvements in CodeBLEU and 92\% higher compilation success, highlighting the effectiveness of dialogue-based fine-tuning.

    % splitting for better perf.
    % \item A paradigm for reasoning-centered supervision: By embedding compiler diagnostics, runtime feedback, and error-correction strategies into training data, our approach enables LLMs to learn not just static mappings but also iterative problem-solving processes.
   \item  We introduce and validate strategies for structuring our dialogue data, demonstrating that different formats are optimal for different tasks. We show that decomposing dialogues into fine-grained \texttt{QS-Pairs} is highly effective for mastering syntactic complexity, while using the full \texttt{Dialogue} trace is superior for preserving semantic correctness, with both formats demonstrating significant improvements over traditional code pairs.
    
\end{itemize}

% To demonstrate the generality of our approach, we apply it to two downstream tasks: \texttt{Fortran→C++}, and \texttt{C++→CUDA}. The resulting datasets comprise 3.65k and 3.92k dialogues, respectively, each containing verified code pairs and associated reasoning traces. Fine-tuning open-weight LLMs on these datasets yields up to a 3.31× improvement in CodeBLEU and a 92\% increase in compilation success rate, illustrating substantial gains in both syntactic accuracy and functional reliability. More broadly, our results establish dialogue-driven dataset generation as a general paradigm for advancing LLM-based code translation across low-resource and domain-specific settings.

To demonstrate the generality of our approach, we apply it to two downstream tasks: \texttt{Fortran→C++} and \texttt{C++→CUDA}. 
% Fine-tuning open-weight models on our generated data yields dramatic improvements in functional correctness, boosting unit test success rates by over 56\% on the C++-to-CUDA task. Crucially, our best fine-tuned 7B model significantly surpasses the performance of larger proprietary systems like Gemini 2.5 Flash on key metrics such as compilation success.
Fine-tuning mid-size open-weight models on our data yields substantial gains across compilation, execution, and unit-test success (e.g., CodeLlama-13B on C++→CUDA improves Unit-Test success rate from 12.5\% to 68.8\%; Fortran→C++ from 42.3\% to 74.1\%), consistent CodeBLEU improvements, and stronger debug-round gains. A dialogue-tuned Qwen2.5-7B outperforms larger proprietary baselines (Gemini 2.5 Flash, Llama 4 Scout 17B) on compilation and execution success, highlighting the cost-efficiency of dialogue-centered supervision.
More broadly, our results demonstrate that dialogue-driven supervision consistently improves functional correctness across multiple open-weight model families.
\section{Background}
\begin{figure*}[!t]
\centering
\includegraphics[width=\textwidth]{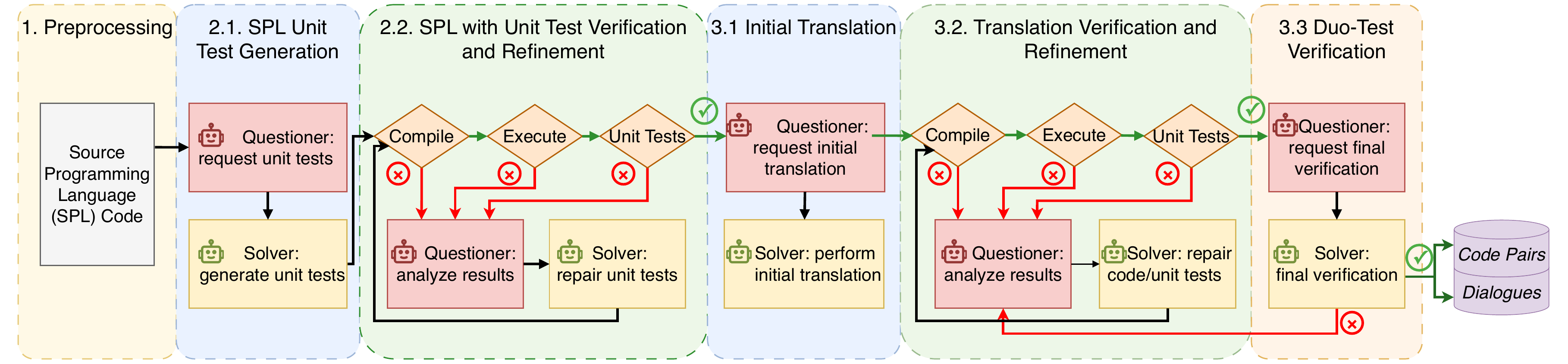} 
\caption{Overview of the Multi-Turn Dialogue Dataset Generation Pipeline. Starting from source programming language (SPL) code, the pipeline generates and refines unit tests through compilation, execution, and validation, then performs test-guided translation with iterative repair based on failure feedback. A final duo test verification stage outputs validated code pairs and dialogues.}
\label{fig:pipeline}
\end{figure*}
\subsection{Code Translation in Software Modernization}
Automated code translation plays a central role in software engineering and high-performance computing~\cite{yang2024exploring, chen2024ompgpt}. Legacy applications written in languages such as Fortran and C need to increasingly interoperate with modern ecosystems, while frameworks such as CUDA, OpenMP, and SYCL have emerged to exploit heterogeneous hardware~\cite{Nichols_2024, ParEval-Repo, mahmud-etal-2025-autoparllm}. Traditional approaches to translation have relied on expert-driven rules, glue code, or intermediate representations, but these methods are costly to maintain and often lack generality across diverse language paradigms~\cite{kadosh2024monocoder}. The rise of large language models has introduced a new direction: using data-driven methods to directly generate target-language code from source-language inputs.

\subsection{Challenges for LLMs in Translation}
Despite progress, LLMs face notable challenges when applied to translation tasks beyond mainstream languages such as Python and Java. First, data scarcity limits generalization in low-resource languages (e.g., Fortran) and in specialized frameworks (e.g., CUDA). Parallel corpora that align source and target languages are rare, particularly when functional equivalence and performance portability must be guaranteed. Second, static source–target pairs only capture the final translation output, omitting the iterative workflows that developers follow when diagnosing compiler errors, validating runtime behavior, and refining translations~\cite{nichols2024hpc, ding2024semcoder, chen2023data}. As a result, fine-tuned models often fail to recover from errors or generate functionally correct code.

\subsection{LLM Agent Systems}
LLM agent frameworks provide a promising way to address these limitations~\cite{mohammadi2025evaluation}. By combining reasoning, planning, memory, and tool integration, agents enable multi-step problem solving that goes beyond single-turn inference. Recent work has demonstrated the value of integrating compilers, execution environments, and feedback loops into translation pipelines, producing data that reflects not only correct outputs but also the reasoning traces behind them~\cite{dearing2024lassi, chen2025pcebench, ding2024semcoder}. However, current agent-based code translation approaches are not able to leverage the intermediate information effectively and automatically.

% systems are particularly well-suited for code translation in low-resource and domain-specific settings, where iterative error correction and external tool use are indispensable for functional correctness.

\section{Approach}
\subsection{Problem and Notation Formulation}

We formulate code translation as a supervised or semi-supervised learning problem. 
In standard settings, the training or fine-tuning dataset consists of paired examples $\{(x_s, x_t)\}$, where $x_s$ is a program in the \textbf{source language} and $x_t$ is its semantically equivalent implementation in the \textbf{target language}. A translation model $f_\theta$ parameterized by $\theta$ then learns to map
\[
x_t = f_\theta(x_s).
\]

As noted earlier, this formulation faces two main challenges. First, reliable parallel corpora $\{(x_s, x_t)\}$ are scarce for low-resource languages and specialized frameworks. Second, conventional datasets capture only the final source–target mapping, omitting the intermediate reasoning required to handle compiler errors, validate runtime behavior, and refine translations. As a result, LLMs trained on static pairs tend to overfit to patterns and struggle to recover from errors.
% This formulation faces two major difficulties in practice. First, for low-resource languages and specialized frameworks (e.g., Fortran, CUDA), there is a scarcity of reliable parallel corpora $\{(x_s, x_t)\}$ for training or adaptation. Second, conventional datasets capture only the final mapping from source to target code, omitting the intermediate reasoning steps that developers and models must follow when handling compiler errors, validating runtime behavior, or refining translations. As a result, LLMs finetuned on static pairs often fail to generalize beyond pattern matching and struggle to recover from translation errors.  

To overcome these limitations, we extend the notion of a dataset to include not only the \textbf{Code Pair} $(x_s, x_t)$ but also the \textbf{Dialogue} that produces it. A dialogue $d$ is a multi-turn sequence of interactions between the Questioner and the Solver,
\[
d = \{ (q_1, s_1), (q_2, s_2), \dots, (q_T, s_T) \},
\]
where $q_i$ is a query and $s_i$ the corresponding response. We refer to each pair $\langle q_i, s_i \rangle$ as a \textbf{Question–Solution Pair}.

This formulation captures not only the final verified source–target pair $(x_s, x_t)$ but also the reasoning trajectory that leads to it. The specific roles of the Questioner and Solver will be detailed in the following subsection.

% In this setup, the Questioner constructs each query $q_i$ by drawing on memory of previous turns, incorporating feedback from external tools, and encoding the relevant problem context and constraints. The Solver, in turn, responds with $s_i$, which may consist of generated code, refinements to address detected issues, or verification results such as test outcomes and error reports. 

% Through this iterative exchange, the dialogue records not only the final verified source–target pair $(x_s, x_t)$ but also the reasoning trajectory that leads to it, providing richer supervision signals than traditional code-only datasets.

\begin{table}[ht]
\centering
\footnotesize
\caption{Roles of the Questioner and Solver in dialogue-based translation.}
\label{tab:qs_roles}
\begin{tabular}{p{0.22\linewidth} p{0.70\linewidth}}
\toprule
\textbf{Role} & \textbf{Main Responsibilities} \\
\midrule
\rowcolor{red!10}
\textbf{Questioner} &
Uses memory of previous turns; incorporates compiler and runtime feedback; specifies context, tags, and constraints; checks source–target consistency. \\
[0.5ex]
\rowcolor{green!10}
\textbf{Solver} &
Generates target-language code; refines or fixes code when errors occur; returns verification results such as tests and error reports. \\
\bottomrule
\end{tabular}
\end{table}

% where $q_i$ is a query (e.g., translation request, error diagnosis) and $a_i$ is the corresponding response (e.g., generated code, corrected snippet, verification result). In this way, each dataset example provides not only a verified source–target program pair but also the reasoning trace leading to it. Training LLMs on such dialogue-augmented data enables models to learn both the *outcome* of translation and the *process* by which correctness is achieved.  

\subsection{The Questioner–Solver Module}
At the core of our approach is the \textit{Questioner–Solver} module, which extends translation beyond static source–target pairs by modeling an interactive dialogue. As illustrated in Figure~\ref{fig:agent}, the module integrates state tracking, external tools, and iterative refinement into the translation process.

The roles of the two parts are summarized in Table~\ref{tab:qs_roles}. The \textbf{Questioner} analyzes the current state and formulates questions by leveraging dialogue memory, which stores past translation attempts, errors, and fixes, and by incorporating feedback from compilers, runtime environments, and scripts. The \textbf{Solver}, in turn, generates target-language translations, produces unit tests, and applies repairs based on the Questioner’s prompts and verification results.

This separation of responsibilities enables explicit modeling and automatic construction of the reasoning process:  the Questioner focuses on analysis and guidance, while the Solver specializes in code generation and correction. Their iterative exchange supports progressive refinement of translations and facilitates the integration of external knowledge sources, ultimately leading to more robust and generalizable models.

% \begin{itemize}
%     \item Introduce the Questioner–Solver module (core novelty).
%     \item Questioner = state analysis, query formulation.
%     \item Solver = translation, unit test generation, repair.
%     \item Memory = stores dialogue history (translation attempts, errors, fixes).
%     \item Tools = compilers, runtime environments, scripts.
%     \item Figure 1: high-level architecture diagram of Questioner–Solver.
%     \item Advantages: reasoning separation, iterative improvement, external knowledge integration
% \end{itemize}

\begin{table*}[!t]
\centering
\caption{Results on \textbf{Fortran2CPP} translation benchmarks. Each cell shows \textit{count (rate)}. 
The best value in each column is highlighted in \textbf{bold}.}
\label{tab:Fortran2CPP_results}
\setlength{\tabcolsep}{4pt}
\resizebox{\textwidth}{!}{
\small
\begin{tabular}{lccc ccc ccc}
\toprule
 &
  \multicolumn{3}{c}{\textbf{CodeLlama-13B-Instruct-hf}} &
  \multicolumn{3}{c}{\textbf{DeepSeek-Coder-6.7B-Instruct}} &
  \multicolumn{3}{c}{\textbf{Qwen2.5-Coder-7B-Instruct}} \\
\cmidrule(lr){2-4} \cmidrule(lr){5-7} \cmidrule(lr){8-10}

\textbf{Setting} &
  \begin{tabular}[c]{@{}c@{}}\textbf{Unit Test}\\\textbf{Success}\end{tabular} &
  \begin{tabular}[c]{@{}c@{}}\textbf{Compilation}\\\textbf{Success}\end{tabular} &
  \begin{tabular}[c]{@{}c@{}}\textbf{Execution}\\\textbf{Success}\end{tabular} &
  \begin{tabular}[c]{@{}c@{}}\textbf{Unit Test}\\\textbf{Success}\end{tabular} &
  \begin{tabular}[c]{@{}c@{}}\textbf{Compilation}\\\textbf{Success}\end{tabular} &
  \begin{tabular}[c]{@{}c@{}}\textbf{Execution}\\\textbf{Success}\end{tabular} &
  \begin{tabular}[c]{@{}c@{}}\textbf{Unit Test}\\\textbf{Success}\end{tabular} &
  \begin{tabular}[c]{@{}c@{}}\textbf{Compilation}\\\textbf{Success}\end{tabular} &
  \begin{tabular}[c]{@{}c@{}}\textbf{Execution}\\\textbf{Success}\end{tabular} \\ 
\midrule

\multicolumn{10}{c}{\textit{Fortran2CPP Code Pair Test (652 tests)}} \\ 
\midrule
\rowcolor{gray!10}
Original      & 276 (42.30\%) & 385 (59.05\%) & 373 (57.21\%) & 480 (73.60\%) & 599 (91.87\%) & 576 (88.34\%) & 444 (68.10\%) & 530 (81.29\%) & 517 (79.29\%) \\
Code Pair     & 449 (68.90\%) & \textbf{597 (91.56\%)} & 570 (87.42\%) & 478 (73.30\%) & \textbf{619 (94.94\%)} & \textbf{602 (92.33\%)} & 411 (63.00\%) & \textbf{604 (92.64\%)} & \textbf{573 (87.88\%)} \\
Dialogue      & \textbf{483 (74.10\%)} & 574 (88.04\%) & 564 (86.50\%) & \textbf{516 (79.10\%)} & 590 (90.49\%) & 582 (89.26\%) & \textbf{491 (75.30\%)} & 570 (87.42\%) & {561 (86.04\%)} \\
QS-Pair       & 442 (67.80\%) & 592 (90.80\%) & \textbf{576 (88.34\%)} & 471 (72.20\%) & 593 (90.95\%) & 576 (88.34\%) & 374 (57.40\%) & 556 (85.28\%) & 523 (80.21\%) \\ 
\midrule

\multicolumn{10}{c}{\textit{HPC-Fortran-Cpp (301 tests)}} \\ 
\midrule
\rowcolor{gray!10}
Original      & 37 (12.29\%) & 115 (38.21\%) & 100 (33.22\%) & \textbf{78 (25.91\%)} & 168 (55.81\%) & 153 (50.83\%) & 70 (23.26\%) & 158 (52.49\%) & 155 (51.50\%) \\
Code Pair     & \textbf{81 (26.91\%)} & \textbf{253 (84.05\%)} & \textbf{238 (79.07\%)} & 74 (24.58\%) & \textbf{254 (84.39\%)} & \textbf{237 (78.74\%)} & \textbf{72 (23.92\%)} & \textbf{248 (82.39\%)} & \textbf{234 (77.74\%)} \\
Dialogue      & 72 (23.92\%) & 204 (67.77\%) & 193 (64.12\%) & 70 (23.26\%) & 229 (76.08\%) & 209 (69.44\%) & 69 (22.92\%) & 225 (74.75\%) & 221 (73.42\%) \\ 
\bottomrule
\end{tabular}}
\end{table*}

\subsection{Multi-Turn Dialogue Dataset Generation Pipeline}

Figure~\ref{fig:pipeline} illustrates our multi-turn dialogue dataset generation pipeline. 
The pipeline integrates the dual-LLM Questioner–Solver module throughout all stages, ensuring that both source–target code pairs and reasoning-rich information are systematically captured and automatically constructed into dialogue data.
Unlike prior works that stop at producing verified code pairs, our design explicitly embeds compiler and runtime feedback into structured multi-turn exchanges, capturing the reasoning traces behind successful translations.

% \subsubsection{Preprocessing}
\noindent
\textbf{1. Preprocessing.}
The pipeline begins with the preprocessing of the source programming language (SPL) code. 
This stage removes comments, filters out dependency-heavy snippets, and enforces a token-length constraint to fit within the LLM context window. 
Only self-contained, executable inputs are retained, ensuring that subsequent testing, translation, and refinement are feasible.

% \subsubsection{Unit Test Generation} 
\noindent
\textbf{2.1 SPL Unit Test Generation.}
After preprocessing, the Questioner requests unit tests in a main function, and the Solver generates them for the source language code. 
These unit tests are embedded directly in the program (e.g., using \texttt{assert} in C++ or conditional checks in Fortran) rather than relying on external frameworks, ensuring that the tests can be executed uniformly across different environments.  

% \subsubsection{Unit Test Verification and Refinement}  
\noindent
\textbf{2.2 SPL with Unit Test Verification and Refinement.}
The generated unit tests are compiled and executed. The Questioner analyzes the results and, in case of any failure, prompts the Solver to repair the tests or refine the code. 
This feedback loop continues until the tests pass. If iterations reach the predefined threshold of seven, the process terminates and discards the sample.
By recording these exchanges, the pipeline captures valuable reasoning traces related to test generation and repair.

% \subsubsection{Initial Translation}  
\noindent
\textbf{3.1 Initial Translation.}
Once unit tests are validated, the Questioner requests an initial target-language translation, and the Solver produces the corresponding code. 
This translation, together with the source code and tests, forms the basis for the iterative refinement stage.  

% \subsubsection{Translation Verification and Refinement}  
\noindent
\textbf{3.2 Translation with Unit Test Verification and Refinement.}
The translated code is compiled and executed against the generated unit tests. The Questioner evaluates the results, and if errors are detected, it instructs the Solver to refine the translation. 
This loop continues until the program passes compilation and test execution or is rejected after repeated failures.  
By capturing intermediate errors and fixes, this stage enriches the dialogue dataset with problem-solving trajectories.  

% \subsubsection{Final Duo-Test Verification}  
\noindent
\textbf{3.3 Duo-Test Verification.}
In this final stage, the Questioner requests a verification step to ensure functional equivalence between the source and translated programs. 
Both codes are compiled and executed, and results are compared. 
If outputs match, the translation and its dialogue history are stored as verified data.  
If mismatches persist, the dialogue is still recorded, including error messages and unsuccessful fixes, making it valuable for training error recovery.  

\subsection{Dataset Output}  
The complete process results in dialogues $d = \{(q_1,s_1), \dots, (q_T,s_T)\}$, where each query $q_i$ is generated by the Questioner and each response $s_i$ is returned by the Solver.  
These dialogues capture not only the final verified source–target program pair $(x_s, x_t)$ but also the reasoning trajectory leading to it.  
This design enables the dataset to serve both as a repository of verified code pairs and as a source of structured supervision for reasoning-centered fine-tuning.

% \subsubsection{Data Preprocessing} Selected input source programming language data will go through a preprocessing step to remove comments and check for external dependencies. Removing comments from seed code before translation ensures that only executable logic is preserved. This helps eliminate unnecessary information, keep the input concise, reduce token consumption, and prevent outdated or irrelevant comments from interfering with code conversion.

% Seed source PL code containing undefined external references will be skipped by this step. If the code relies on external dependencies that are not provided in the snippet, the translation to C++ may be incomplete or incorrect.

% We also apply two filtering criteria: limiting token count to less than a threshold (e.g. 600) to fit within LLM's input context window and including only executable code to facilitate unit testing. 

% \subsubsection{}

\subsection{Dataset Construction}
Compared with code-pair data, a major advantage of dialogue data lies in its flexibility for constructing downstream tasks. By default, we adopt Dialogue-Level Splitting for fine-tuning, though the data can also be divided at the Question–Solution Pair level to mitigate potential overfitting.

\noindent
\textbf{Dialogue-Level Splitting}:
Our dataset consists of multi-turn dialogues generated by the dual-agent Questioner–Solver system, where each dialogue captures a full translation workflow—initial translation, unit test generation, error diagnosis, iterative correction, and final verification. To preserve the natural flow of interactions, we split the data at the dialogue level. Each dialogue is assigned entirely to the training, validation, or test set, ensuring that no portion of a conversation appears in multiple splits. This prevents context leakage and allows evaluation on truly unseen translation scenarios.

\noindent
\textbf{Question–Solution Pair Splitting}:
Each multi-turn dialogue can be further decomposed into cumulative Question–Solution pairs for model fine-tuning. This splitting strategy emphasizes individual interactions and applies random sampling across pairs. It preserves the iterative refinement and error-correction context essential for model learning while mitigating potential overfitting. We posit that such finer-grained and randomized splitting enhances generalization by exposing the model to a broader and more diverse set of interaction patterns—a hypothesis that we further analyze in the discussion section.
% \subsection{Generalization discussion}

\section{Experiments}
\label{sec-experiments}

% In this section, we present the experimental setup and results of using Fortran2CPP fine-tuning selected LLMs for Fortran to C++ translation. %Weevaluating the performance of large language models (LLMs) in translating Fortran code to C++. 

This section validates our proposed dialogue-based data generation pipeline through a series of experiments on two critical, low-resource code translation tasks. The first, \textbf{Fortran-to-C++ (Fortran2CPP)}, addresses the challenge of updating legacy scientific code. The second, \textbf{C++-to-CUDA (CPP2CUDA)}, focuses on modernizing code for heterogeneous parallelization and GPU kernel synthesis. Our analysis is structured to answer three core research questions that systematically evaluate our method's performance, advantages, and competitiveness.

% To validate our dialogue-based data generation pipeline, this section details experiments on two critical, low-resource tasks: modernizing legacy Fortran code to C++ (\textbf{Fortran2CPP}) and parallelizing C++ for GPU execution via CUDA (\textbf{CPP2CUDA}). Our analysis is structured to answer three core research questions that systematically evaluate our method's performance, advantages, and competitiveness.

%==============================================

\subsection{Datasets}
\label{sec:datasets}

The datasets for these two tasks were generated via our pipeline, which queried the Llama-3.3-70B-Instruct~\citep{meta2024llama3} and Llama-4-Scout-17B-16E-Instruct~\citep{meta2024llama4scout} models through the Google Vertex AI API~\citep{google2024vertexai}. The resulting corpora span a diverse range of language constructs, including array operations, loop nesting patterns, pointer arithmetic, and parallelism primitives (e.g., OpenMP directives in Fortran, CUDA kernel launches in C++). To ensure a rigorous evaluation, we split all generated data by source file index ranges into distinct training and testing partitions, guaranteeing no overlap.

\begin{figure*}[!tb]
\centering
\includegraphics[width=\textwidth]{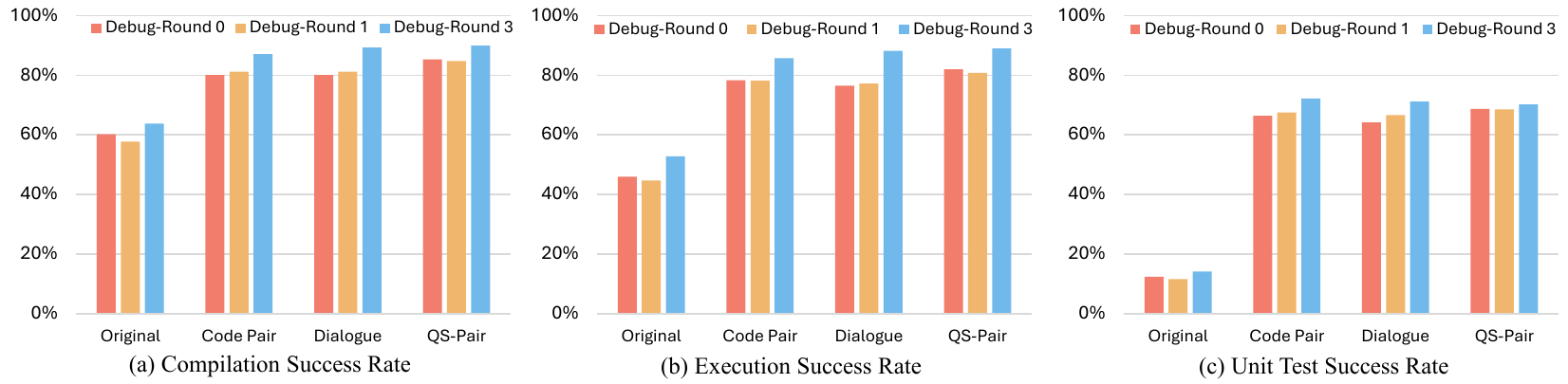} 
\caption{Impact of Fine-tuning and Debug Rounds on C++ to CUDA Translation Success (CodeLlama-13B).}
\vspace{-5pt}
\label{fig:cuda2cpp_debug_round}
\end{figure*}

\subsection{Results and Analysis}
\label{sec:results}

\subsubsection{Fine-tuning Datasets}

\paragraph{Fortran2CPP Translation.}
We sampled Fortran programs from the CodeParrot GitHub-Code dataset~\citep{codeparrot}, a collection of over 142,000 files from open-source scientific repositories. Our pipeline processed a contiguous block of 40,000 samples (indices 70,000--110,000) and produced 3,652 validated Fortran-to-C++ pairs, each confirmed to compile and execute successfully. The first 3,000 of these pairs formed the training set.

\paragraph{CPP2CUDA Translation.}
We built upon the CodeRosetta corpus~\citep{coderosetta2024}, a large synthetic collection for unsupervised parallel programming translation. From its 6,033 C++--CUDA pairs, we retained only the C++ sources and re-translated them, yielding 3,394 verified translations. The first 3,000 pairs were used for training. Each training entry includes the source C++ function, the generated CUDA kernel with a runnable test unit, and multi-round dialogue logs that record compiler errors and solver fixes.

\paragraph{Data Formats.}
To analyze the impact of supervision structure, we prepared three dataset variants: \textbf{Code Pair}, containing direct source-target translations; \textbf{Dialogue}, comprising full multi-turn reasoning traces with compiler and runtime feedback; and \textbf{Question-Solution (QS) Pair}, consisting of atomic QS exchanges extracted from dialogues, which yielded 14,182 pairs for Fortran2CPP and 12,770 for CPP2CUDA.

\subsubsection{Evaluation Benchmarks}

\paragraph{Fortran2CPP Evaluation Sets.}
We used the remaining 652 pairs from our generated Fortran-to-C++ corpus as the \texttt{Fortran2CPP Code Pair Test} set. To assess broader generalization, we also adopted the HPC-Fortran-Cpp dataset~\citep{lei2023creating}, which contains 315 manually curated OpenMP Fortran/C++ pairs. After filtering for sources under 4,000 tokens, 301 pairs were retained for evaluation. As only 116 of the 301 Fortran source programs in the HPC-Fortran-Cpp set are executable (the remainder require external dependencies or lack runnable entry points), the Unit Test Success rate is consistently low for this benchmark. 

\paragraph{CPP2CUDA Evaluation Sets.}
Two complementary test sets were derived from the CodeRosetta source corpus, disjoint from the training indices. The first is the  \texttt{CPP2CUDA Code Pair Test with Checksum}, containing the last 394 pairs evaluated under strict verification that enforces bit-exact checksum consistency between CPU and GPU outputs. The second is the \texttt{CPP2CUDA Code Pair Test}, an additional 528 pairs drawn from the remaining 800 samples that preceded the test subset. Both test sets were generated using the same pipeline configuration.

\subsection{Experiment Setup}

\begin{table*}[!tbh]
\centering
\caption{Results on \textbf{CPP2CUDA} translation benchmarks. Each cell shows \textit{count (rate)}. 
The best value in each column is highlighted in \textbf{bold}.}
\label{tab:CPP2CUDA_unit_comp_exec}
\setlength{\tabcolsep}{4pt}
\resizebox{\textwidth}{!}{
\small
\begin{tabular}{lccc ccc ccc}
\toprule
 &
  \multicolumn{3}{c}{\textbf{CodeLlama-13B-Instruct-hf}} &
  \multicolumn{3}{c}{\textbf{DeepSeek-Coder-6.7B-Instruct}} &
  \multicolumn{3}{c}{\textbf{Qwen2.5-Coder-7B-Instruct}} \\
\cmidrule(lr){2-4} \cmidrule(lr){5-7} \cmidrule(lr){8-10}

\textbf{Setting} &
  \begin{tabular}[c]{@{}c@{}}\textbf{Unit Test}\\\textbf{Success}\end{tabular} &
  \begin{tabular}[c]{@{}c@{}}\textbf{Compilation}\\\textbf{Success}\end{tabular} &
  \begin{tabular}[c]{@{}c@{}}\textbf{Execution}\\\textbf{Success}\end{tabular} &
  \begin{tabular}[c]{@{}c@{}}\textbf{Unit Test}\\\textbf{Success}\end{tabular} &
  \begin{tabular}[c]{@{}c@{}}\textbf{Compilation}\\\textbf{Success}\end{tabular} &
  \begin{tabular}[c]{@{}c@{}}\textbf{Execution}\\\textbf{Success}\end{tabular} &
  \begin{tabular}[c]{@{}c@{}}\textbf{Unit Test}\\\textbf{Success}\end{tabular} &
  \begin{tabular}[c]{@{}c@{}}\textbf{Compilation}\\\textbf{Success}\end{tabular} &
  \begin{tabular}[c]{@{}c@{}}\textbf{Execution}\\\textbf{Success}\end{tabular} \\ 
\midrule

\multicolumn{10}{c}{\textit{CPP2CUDA Code Pair Test (528 tests)}} \\ 
\midrule
\rowcolor{gray!10}
Original      & 66 (12.50\%) & 318 (60.23\%) & 243 (46.02\%) & 306 (58.00\%) & 418 (79.17\%) & 395 (74.81\%) & 260 (49.20\%) & 349 (66.10\%) & 316 (59.85\%) \\
Code Pair     & 351 (66.50\%) & 423 (80.11\%) & 414 (78.41\%) & 368 (69.70\%) & 440 (83.33\%) & 428 (81.06\%) & 342 (64.80\%) & 417 (78.98\%) & 406 (76.89\%) \\
Dialogue      & 339 (64.20\%) & 423 (80.11\%) & 404 (76.52\%) & 354 (67.00\%) & 435 (82.39\%) & 423 (80.11\%) & 339 (64.20\%) & 436 (82.58\%) & 421 (79.73\%) \\
QS-Pair       & \textbf{363 (68.80\%)} & \textbf{451 (85.42\%)} & \textbf{433 (82.01\%)} & \textbf{377 (71.40\%)} & \textbf{453 (85.80\%)} & \textbf{443 (83.90\%)} & \textbf{358 (67.80\%)} & \textbf{458 (86.74\%)} & \textbf{444 (84.09\%)} \\ 
\midrule

\multicolumn{10}{c}{\textit{CPP2CUDA Code Pair Test with Checksum (394 tests)}} \\ 
\midrule
\rowcolor{gray!10}
Original      & 26 (6.60\%) & 143 (36.29\%) & 136 (34.52\%) & 230 (58.40\%) & 267 (67.77\%) & 260 (65.99\%) & 184 (46.70\%) & 227 (57.61\%) & 222 (56.35\%) \\
Code Pair     & \textbf{299 (75.90\%)} & \textbf{332 (84.26\%)} & \textbf{328 (83.25\%)} & \textbf{305 (77.40\%)} & \textbf{335 (85.03\%)} & \textbf{329 (83.50\%)} & \textbf{281 (71.30\%)} & \textbf{332 (84.26\%)} & \textbf{328 (83.25\%)} \\
Dialogue      & 275 (69.80\%) & 304 (77.16\%) & 300 (76.14\%) & 286 (72.60\%) & 320 (81.22\%) & 315 (79.95\%) & 273 (69.30\%) & 316 (80.20\%) & 313 (79.44\%) \\
QS-Pair       & 288 (73.10\%) & 329 (83.50\%) & 323 (81.98\%) & 289 (73.40\%) & 326 (82.74\%) & 321 (81.47\%) & 277 (70.30\%) & {331 (84.01\%)} & {327 (82.99\%)} \\ 
\bottomrule
\end{tabular}}
\end{table*}

\noindent

\paragraph{Models.}
For our fine-tuning experiments, we selected a range of strong, publicly available code-oriented language models (details in Table~\ref{tab:selected-models} in Appendix). These include {CodeLlama-13B-Instruct-hf}~\citep{roziere2023codellama}, {DeepSeek-Coder-6.7B-Instruct}~\citep{guo2024deepseekcoder}, and {Qwen2.5-Coder-7B-Instruct}~\citep{qwen2024qwen2}, which represent different architectures and training data sizes. To benchmark the performance of our fine-tuned models, we compare them against two state-of-the-art proprietary models: {Gemini 2.5 Flash}~\citep{google2024gemini} and {Llama 4 Scout 17B}~\citep{meta2024llama4scout}.

\paragraph{Evaluation Metrics.}
We assess the quality of the translated code using four complementary metrics that evaluate syntactic, runtime, functional, and structural correctness:
\begin{itemize}[nosep]
    \item \textbf{Compilation Success Rate:} This metric measures the syntactic validity of the generated code. A translation is considered successful if it compiles without any errors using standard compilers (e.g., GCC for C++, NVCC for CUDA).
    \item \textbf{Execution Success Rate:} For code that compiles successfully, this metric assesses its runtime stability. A translation passes if the compiled executable runs to completion without encountering any runtime errors, such as segmentation faults.
    \item \textbf{Unit Test Success Rate:} This is our primary and most stringent metric for functional correctness. It evaluates whether the translated code produces the correct output, as verified by a set of unit tests.
    \item \textbf{CodeBLEU Score:} This metric evaluates the quality of translated code by measuring its similarity to a reference solution. It extends the standard BLEU score by incorporating code-specific features like n-gram, abstract syntax tree (AST), and dataflow matching.
\end{itemize}

\noindent
\textbf{Implementation Details:}            
All experiments were conducted on an NVIDIA H200 GPU (140 GB) using LlamaFactory~\cite{zheng2024llamafactory} for model training. We fine-tuned each model with LoRA~\cite{hu2021loralowrankadaptationlarge} under the supervised fine-tuning (SFT) stage on two datasets: one split at the conversation level and one at the prompt-response pair level, each containing 3,000 samples. The Dialogue-Model was trained with a context length of 8,192, while the Pair-Model used 4,096. LoRA was applied to all target modules with rank $= 8$, $\alpha= 16$. Training used a per-device batch size $=1$, gradient accumulation = 8, learning rate = $1 \times 10^{-4}$, a cosine learning-rate scheduler with warm-up ratio $= 0.1$, and ran for 3 epochs with fp16 precision enabled. Each fine-tuning experiment takes $\sim$4–6 GPU hours on a single GPU.

\noindent
For inference, we merged the LoRA adapters with the base model weights and evaluated the test split with a temperature of 0.2 on vLLM~\cite{kwon2023efficient}. This setup provides a comprehensive evaluation of LLM performance for Fortran-to-C++ translation across our datasets and metrics.

\begin{figure*}[!tb]
\centering
\includegraphics[width=\textwidth]{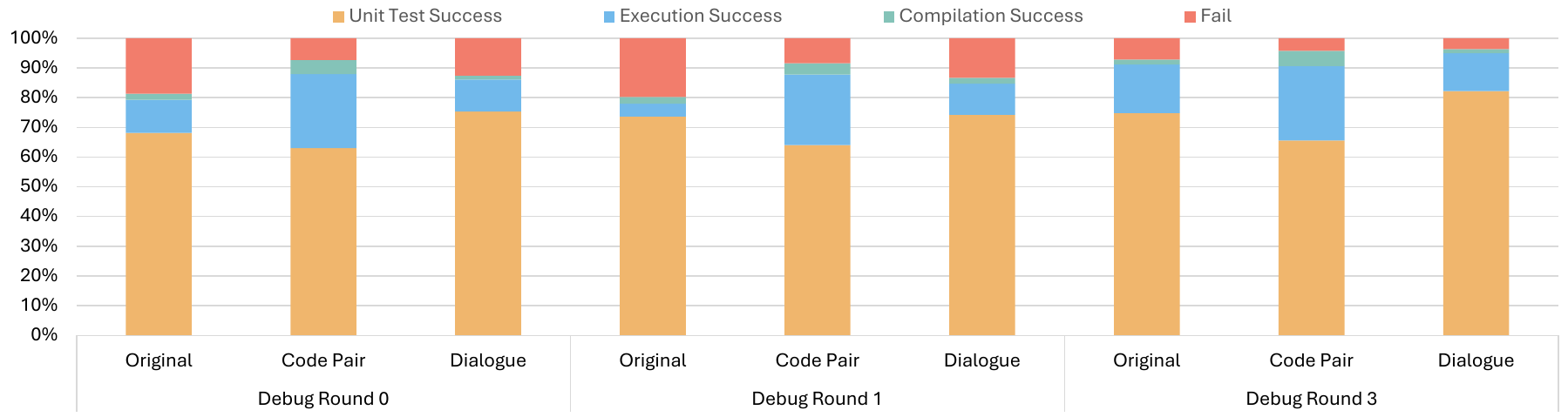} 
\caption{Impact of Fine-tuning and Debug Rounds on Fortran to C++ Translation Success (Qwen2.5-Coder-7B).}
\label{fig:f2c_debug_round}
\end{figure*}

\paragraph{RQ1: Can the proposed pipeline effectively generate high-quality data that supports LLMs in code translation tasks?} 
To answer this question, we evaluate the effectiveness of our dataset by fine-tuning multiple models and comparing their performance against their original counterparts using compilation success, execution success, and unit-test pass rates. 
We first conduct a coverage audit to assess the rigor and quality of the generated unit tests. Using the \texttt{gcov} profiling tool, we observe that the generated test suites achieve an average line coverage of 87.3\% and a branch coverage of 83.8\%, indicating that the pipeline produces comprehensive and non-trivial tests that exercise diverse execution paths.
As shown in Table~\ref{tab:CPP2CUDA_unit_comp_exec}, the baseline CodeLlama-13B model achieves a Unit Test Success Rate of just 12.50\% on the main CPP2CUDA test set. After fine-tuning with our QS-Pair data, its success rate soars to 68.80\% (+56.3\%). This trend holds consistently for the Fortran2CPP task (Table~\ref{tab:Fortran2CPP_results}), where CodeLlama-13B improves from 42.3\% to 74.1\%, DeepSeek-6.7B from 73.6\% to 79.1\%, and Qwen2.5-7B from 68.1\% to 75.3\%. These upward trends align with CodeBLEU gains (Table \ref{tab:codebleu}), confirming that our datasets enhance both syntactic validity and functional correctness. This strongly demonstrates that our pipeline produces large-scale, high-quality translation supervision that reliably benefits multiple model families.

\begin{table*}[tbh]
\centering
\caption{Comparison of model performance on \textbf{CPP2CUDA} and \textbf{Fortran2CPP} code pair tests. 
The results show compilation, execution, and overall unit test success rates across different fine-tuned models.}
\label{tab:sota_compare}
\resizebox{\textwidth}{!}{
\begin{tabular}{llccc|ccc}
\toprule
\multirow{2}{*}{\textbf{Model}} & \multirow{2}{*}{\textbf{Size}} &
\multicolumn{3}{c|}{\textbf{CPP2CUDA Code Pair Test}} &
\multicolumn{3}{c}{\textbf{Fortran2CPP Code Pair Test}} \\ \cmidrule(lr){3-5} \cmidrule(lr){6-8}
 & & \textbf{Compilation} & \textbf{Execution} & \textbf{Unit Test} 
 & \textbf{Compilation} & \textbf{Execution} & \textbf{Unit Test} \\ 
\midrule
Gemini 2.5 Flash & Undisclosed & 79.17\% & 73.86\% & 63.30\% & 91.87\% & 90.64\% & \textbf{87.30}\% \\
Llama 4 Scout 17B & 17B & 78.98\% & 75.57\% & 64.80\% & 91.10\% & 89.88\% & 85.60\% \\
\midrule
Qwen2.5 Coder 7B (Dialogue, D-round=3) & 7B & \textbf{93.18\%} & \textbf{92.05\%} & \textbf{72.70\%} & \textbf{96.32\%} & \textbf{94.94\%} & {82.20\%} \\ 
\bottomrule
\end{tabular}
}
\end{table*}

\begin{table*}[!t]
\centering
\caption{
CodeBLEU scores on Fortran2CPP and CPP2CUDA tasks across different fine-tuning settings.
}
\label{tab:codebleu}
\setlength{\tabcolsep}{3pt}        % 
\renewcommand{\arraystretch}{1} % 
\resizebox{\textwidth}{!}{
% \small
\begin{tabular}{lcccc cccc cccc cc}
\toprule
& \multicolumn{4}{c}{\textbf{CodeLlama-13B}} &
  \multicolumn{4}{c}{\textbf{DeepSeek-Coder-6.7B}} &
  \multicolumn{4}{c}{\textbf{Qwen2.5-Coder-7B}} &
  \multicolumn{2}{c}{\textbf{Large Baselines}} \\
\cmidrule(lr){2-5} \cmidrule(lr){6-9} \cmidrule(lr){10-13} \cmidrule(lr){14-15}
\textbf{Task} 
& Original & Code Pair & Dialogue & QS-Pair 
& Original & Code Pair & Dialogue & QS-Pair  
& Original & Code Pair & Dialogue & QS-Pair 
& Gemini & Llama4 \\
\midrule
Fortran2CPP 
& 0.565 & \textbf{0.596} & 0.594 & 0.591 
& 0.542 & \textbf{0.597} & 0.592 & 0.593 
& 0.556 & 0.581 & \textbf{0.585} & 0.571 
& 0.593 & 0.593 \\
CPP2CUDA 
& 0.559 & 0.628 & 0.623 & \textbf{0.629} 
& 0.574 & 0.626 & 0.626 & \textbf{0.629} 
& 0.581 & \textbf{0.619} & 0.618 & 0.616 
& 0.629 & 0.630
 \\
\bottomrule
\end{tabular}
}
\end{table*}

% \paragraph{RQ2: Compared with conventional Code Pair datasets, does incorporating compiler and runtime feedback as well as reasoning traces within multi-round dialogues improve translation accuracy and robustness?}
% Our results indicate that dialogue-centric formats (`Dialogue` and `QS-Pair`) are highly competitive and often superior to traditional `Code Pair` fine-tuning, exhibiting complementary strengths across different tasks. For the Fortran2CPP task, which requires preserving semantic integrity, the full `Dialogue` format consistently yields the highest Unit Test Success Rate across all three models (Table~\ref{tab:Fortran2CPP_results}), reaching 74.10\% for CodeLlama-13B. Conversely, for the CPP2CUDA task, which demands precise adherence to syntax and compilation rules, the more granular `QS-Pair` format is particularly effective, delivering the best Unit Test performance for all three models on the main test set (Table~\ref{tab:CPP2CUDA_unit_comp_exec}). Furthermore, the value of the iterative refinement process captured in dialogues is evident from the impact of debug rounds (Figures 3 and 4, not shown in current draft). Increasing the number of rounds monotonically improves compilation, execution, and success rates, demonstrating that the dialogue-driven feedback loop teaches models how to recover from failures, not just perform a one-shot translation.

\paragraph{RQ2: Compared with conventional Code Pair datasets, does incorporating compiler and runtime feedback as well as reasoning traces within multi-round dialogues improve translation accuracy and robustness?}

Dialogue-centric formats (Dialogue and QS-Pair) are highly competitive and often superior to conventional Code Pair fine-tuning, with complementary strengths across tasks. On \textbf{Fortran2CPP}, full \textit{Dialogue} achieves the best functional correctness (e.g., 74.10\% Unit Test Success for CodeLlama-13B; Table~\ref{tab:Fortran2CPP_results}), as its rich reasoning context helps models master algorithmic details when syntactic correctness is already high. On the harder \textbf{CPP2CUDA} task, \textit{QS-Pair} performs best (Table~\ref{tab:CPP2CUDA_unit_comp_exec}): decomposing 3k dialogues into $>$12k shuffled turns provides more optimization steps within the same three epochs, accelerating convergence to stable compilation/execution patterns.

To further probe these benefits, we introduce a \textit{debug round} mechanism that returns compilation errors to the model for iterative repair. Figure~\ref{fig:f2c_debug_round} shows that Code-Pair achieves the highest \emph{compilation} success each round, yet its \emph{unit-test} success remains below Dialogue. The Dialogue-tuned model improves across all metrics with each round and by D-Round~3 surpasses both Code-Pair and QS-Pair in compilation, execution, and unit-test success, demonstrating that feedback-rich dialogues teach models to \emph{correct and refine} translations rather than merely generate compilable templates. Figure~\ref{fig:cuda2cpp_debug_round} confirms the same trend on CPP2CUDA, with Dialogue exhibiting the steepest improvements. By matching supervision granularity to task difficulty—full \textit{Dialogue} for semantic tasks, fine-grained \textit{QS-Pair} for syntactically demanding ones—dialogue-based supervision yields code that is both syntactically sound and functionally correct.

\paragraph{RQ3: Can small open-weight models fine-tuned on dialogue-centric data achieve competitiveness with state-of-the-art proprietary systems?}
Yes. Our fine-tuned, mid-size open-weight models demonstrate the ability to match or even surpass the performance of larger systems. As detailed in Table~\ref{tab:sota_compare}, our Dialogue fine-tuned Qwen2.5-7B models with Debug rounds 3 significantly outperform both Gemini 2.5 Flash and Llama 4 Scout 17B in Compilation, Execution, and Unit Test Success rates. A similar trend is observed on the Fortran2CPP task, where our model's Execution Success Rate of 94.94\% exceeds that of the larger baselines. This parameter-efficient competitiveness is further supported by the CodeBLEU scores in Table~\ref{tab:codebleu}, where our best models are on par with or exceed the large baselines. These findings highlight that our data generation method is a cost-effective pathway to achieving state-of-the-art performance.

% \FloatBarrier 
\section{Conclusion}
This paper addresses the challenges of low-resource code translation by introducing an LLM-based approach featuring a novel Questioner--Solver module.
Our framework automatically generates multi-turn dialogue datasets that capture reasoning traces, compiler feedback, and error-correction strategies, providing richer supervision than traditional code-pair datasets.
Experiments on Fortran\(\to\)C++ and C++\(\to\)CUDA demonstrate substantial improvements in compilation, execution, and unit-test success rates, with a fine-tuned 7B model surpassing larger proprietary baselines.
These results highlight the potential of dialogue-based supervision to significantly advance legacy code modernization.
Future efforts will focus on expanding language coverage, incorporating performance-aware evaluation, and exploring curriculum-based training strategies.

% \subsection{Appendices}

% Use \verb|\appendix| before any appendix section to switch the section numbering over to letters. See Appendix~\ref{sec:appendix} for an example.

% \section{Bib\TeX{} Files}
% \label{sec:bibtex}

% Unicode cannot be used in Bib\TeX{} entries, and some ways of typing special characters can disrupt Bib\TeX's alphabetization. The recommended way of typing special characters is shown in Table~\ref{tab:accents}.

% Please ensure that Bib\TeX{} records contain DOIs or URLs when possible, and for all the ACL materials that you reference.
% Use the \verb|doi| field for DOIs and the \verb|url| field for URLs.
% If a Bib\TeX{} entry has a URL or DOI field, the paper title in the references section will appear as a hyperlink to the paper, using the hyperref \LaTeX{} package.

% \section*{Limitations}
% Our framework, though effective, relies on simplified test environments and currently targets only Fortran→C++ and C++→CUDA. Dialogue quality depends on base model reasoning, and evaluation focuses mainly on compilability and correctness rather than performance. Future work will broaden language coverage, include performance checks, and improve efficiency.

\section*{Limitations}
Our framework, though effective, has several limitations. 
The current framework focuses primarily on functional correctness, a challenging and necessary first step, measured by compilation, execution, and unit-test success.  
Performance portability metrics such as memory access efficiency and kernel execution time are not yet integrated into our automated feedback loop. 
Future work will broaden language coverage, incorporate performance-aware evaluation, and improve efficiency.

\section*{Acknowledgments}
This work was prepared in part by Lawrence Livermore National Laboratory under Contract DE-AC52-07NA27344 (LLNL-CONF-2018168). It was also supported by the U.S. Department of Energy, Office of Science, under Contract DE-AC02-06CH11357, and by the National Science Foundation under Grant No. 2211982. The United States Government retains, and the publisher, by accepting the article for publication, acknowledges that the United States Government retains a non-exclusive, paid-up, irrevocable, worldwide license to publish or reproduce the published form of this manuscript, or to allow others to do so, for United States Government purposes.

% \section*{Acknowledgments}

% This document has been adapted
% by Steven Bethard, Ryan Cotterell and Rui Yan
% from the instructions for earlier ACL and NAACL proceedings, including those for
% ACL 2019 by Douwe Kiela and Ivan Vuli\'{c},
% NAACL 2019 by Stephanie Lukin and Alla Roskovskaya,
% ACL 2018 by Shay Cohen, Kevin Gimpel, and Wei Lu,
% NAACL 2018 by Margaret Mitchell and Stephanie Lukin,
% Bib\TeX{} suggestions for (NA)ACL 2017/2018 from Jason Eisner,
% ACL 2017 by Dan Gildea and Min-Yen Kan,
% NAACL 2017 by Margaret Mitchell,
% ACL 2012 by Maggie Li and Michael White,
% ACL 2010 by Jing-Shin Chang and Philipp Koehn,
% ACL 2008 by Johanna D. Moore, Simone Teufel, James Allan, and Sadaoki Furui,
% ACL 2005 by Hwee Tou Ng and Kemal Oflazer,
% ACL 2002 by Eugene Charniak and Dekang Lin,
% and earlier ACL and EACL formats written by several people, including
% John Chen, Henry S. Thompson and Donald Walker.
% Additional elements were taken from the formatting instructions of the \emph{International Joint Conference on Artificial Intelligence} and the \emph{Conference on Computer Vision and Pattern Recognition}.

\bibliography{main}

\appendix
% \section{Data Example}
\section{LLM Selection}

Table~\ref{tab:selected-models} summarizes the representative code-oriented language models used in our evaluation. 
We select these models to cover a range of architectures, parameter scales, and training corpora, while prioritizing reproducibility and tunability.

\begin{table*}[t]
\footnotesize
\centering
\caption{Selected Code-Oriented Language Models for Evaluation}
\label{tab:selected-models}
\begin{adjustbox}{max width=0.95\textwidth}
\begin{tabular}{l|ccc|cc}
\toprule
\textbf{Specification} & \textbf{DeepSeek-Coder} & \textbf{CodeLlama} & \textbf{Qwen2.5-Coder} & \textbf{Gemini 2.5 Flash} & \textbf{Llama 4 Scout} \\
\midrule
Parameters & 6.7B & 13B & 7B & Unknown & 17B \\
Training Data & 2T tokens & 500B tokens & 3T tokens & Unknown & Unknown \\
Context Window & 16K & 100K & 128K & 2M & Unknown \\
Open Weights & Yes & Yes & Yes & No & Yes \\
Developer & DeepSeek AI & Meta & Alibaba Cloud & Google & Meta \\
\bottomrule
\end{tabular}
\end{adjustbox}
\end{table*}

\paragraph{On the Choice of Open-Weight Models.}
While frontier models (e.g., GPT-family systems) can provide additional insight into the upper bound of data quality and iterative refinement, our pipeline is intentionally designed around reproducible open-weight models. This design choice is motivated by both practical and methodological considerations. 

First, the full dataset generation and fine-tuning pipeline involves large-scale iterative execution, compilation, and unit-test feedback loops, which are subject to cost and API throughput constraints when using proprietary models. More importantly, open-weight models allow the community to fully reproduce, re-run, and extend our experiments without requiring access to closed or proprietary systems. In addition, open-weight models support supervised and continued fine-tuning, which is not feasible for most API-based frontier models and is essential for evaluating the downstream impact of the generated supervision.

\paragraph{Robustness Across Model Choices.}
To assess the robustness of our model selection, we conducted preliminary experiments on a randomly sampled subset of 100 translation tasks using several representative open-source models. For each model, we report the unit-test success rate, the average number of feedback iterations required, and the end-to-end runtime. These results are summarized in Table~\ref{tab:appendix_model_robustness}. Overall, we observe consistent trends across models, suggesting that the effectiveness of the proposed pipeline is not tied to a specific backbone choice but generalizes across open-weight model families.

\begin{table*}[t]
\centering
\caption{Preliminary robustness analysis across open-weight models on a 100-sample subset. We report unit-test success counts, temperature settings, and approximate end-to-end runtime.}
\begin{tabular}{lccc}
\toprule
\textbf{Model} & \textbf{Temp.} & \textbf{Success} & \textbf{Approx. Runtime} \\
\midrule
DeepSeek-Coder-V2-Lite-Instruct & 0.1 & 4  & $\sim$1h 10m \\
DeepSeek-Coder-V2-Lite-Instruct & 0.3 & 4  & $\sim$25m \\
DeepSeek-Coder-V2-Lite-Instruct & 0.5 & 3  & $\sim$25m \\
DeepSeek-Coder-V2-Lite-Instruct & 0.7 & 6  & $\sim$25m \\
\midrule
DeepSeek-R1-Distill-Qwen-32B & 0.1 & 16 & $\sim$8h \\
DeepSeek-R1-Distill-Qwen-32B & 0.6 & 10 & $\sim$6.5h \\
\midrule
DeepSeek-R1-Distill-Llama-70B (Q4\_K) & 0.6 & 12 & $\sim$4.5h \\
deepseek-r1:70b-llama-distill-q8\_0 & 0.1 & 8  & $\sim$5h \\
deepseek-r1:70b-llama-distill-q8\_0 & 0.6 & 5  & $\sim$2h \\
deepseek-r1:70b-llama-distill-q8\_0 & 0.7 & 9  & $\sim$5h \\
\midrule
Llama-3.3-70B-Instruct (Q4\_K\_M) & 0.7 & 15 & $\sim$3.5h \\
Llama-3.3-70B-Instruct (Q6\_K) & 0.1 & 18 & $\sim$5.4h \\
Llama-3.3-70B-Instruct (Q6\_K) & 0.1 & 12/64\textsuperscript{\dag} & -- \\
Llama-3.3-70B-Instruct (Q6\_K) & 0.3 & 11 & $\sim$3.5h \\
Llama-3.3-70B-Instruct (Q6\_K) & 0.6 & 18 & $\sim$5.5h \\
\midrule
Llama-3.3-70B-Instruct (API) & 0.1 & N/A\textsuperscript{\ddag} & $\sim$1.5h \\
Llama-3.3-70B-Instruct (API) & 0.6 & N/A\textsuperscript{\ddag} & $\sim$1.5h \\
Llama-3.3-70B-Instruct (API) & 0.7 & N/A\textsuperscript{\ddag} & $\sim$1.5h \\
\midrule
DeepSeek-R1-671B (1.73-bit) & 0.6 & N/A\textsuperscript{\S} & $\sim$5h \\
\bottomrule
\end{tabular}
\begin{flushleft}
\footnotesize
\textsuperscript{\dag}Run on a 64-sample subset only. \textsuperscript{\ddag}Pipeline terminated before unit-test evaluation due to API errors. \textsuperscript{\S}Model produced malformed outputs; no valid translations obtained.
\end{flushleft}
\label{tab:appendix_model_robustness}
\end{table*}

\section{Discussion: Toward Fine-Grained Dialogue Utilization}

Beyond the demonstrated benefits of dialogue-based supervision, we further explore
how to more effectively exploit the internal structure of these dialogues to construct
richer training signals. In our current datasets, each full dialogue can be decomposed into
multiple \textit{Question–Solution (QS)} pairs, resulting in a substantially enlarged
fine-tuning corpus (12,770 pairs for CPP2CUDA and 14,182 for Fortran2CPP; see Section 4.1).

Empirically, QS-Pair supervision achieves the best overall performance on the CPP2CUDA
task (Table~\ref{tab:CPP2CUDA_unit_comp_exec}), confirming that fine-grained “question–answer”
signals help models learn robust compiler- and runtime-aware translation patterns.
Fortran2CPP, on the other hand, benefits more from complete dialogue reasoning traces,
suggesting that full-context reasoning captures algorithmic and semantic consistency.

\paragraph{Qualitative Analysis of Error Recovery.} 
We observed that the multi-turn nature of our dataset allows models to learn specific recovery strategies. For instance, in the CPP2CUDA task, common failures often involve incorrect thread indexing or memory synchronization. Our Questioner-Solver logs capture how models transition from a "Segment Fault" error to a corrected kernel by adjusting block dimensions or adding necessary \texttt{\_\_syncthreads()} calls. This progressive refinement is a key differentiator from static code-pair datasets, which only present the final, error-free version.

These findings motivate several directions for future refinement:
\begin{itemize}
    \item \textbf{Round-aware curriculum.} Sample dialogue turns progressively by round depth or error type to form a structured learning schedule.
    \item \textbf{Feedback-tagged prompts.} Explicitly label compiler errors (syntax, linkage, parallel primitives, memory) to guide targeted repair learning.
    \item \textbf{Dual-view mixing.} Combine Dialogue and QS-Pair views within each epoch, balancing high-level reasoning with dense supervision.
\end{itemize}

Overall, these results suggest that multi-round dialogue data not only enhances current
fine-tuning, but also provide a foundation for more systematic curriculum and retrieval-augmented
training strategies in code translation.

\section{Ablation Study on Feedback Signals}

To understand the contribution of different interaction signals, we first analyze the distribution of debugging feedback in our dialogue datasets. Execution-related feedback is the most prevalent signal, appearing in 65.0\% of Fortran2CPP and 44.2\% of CPP2CUDA dialogues, and accounting for a substantial portion of the reasoning traces (6,908 and 4,816 messages, respectively). 
In contrast, compile-error feedback is considerably less frequent, especially for CPP2CUDA (6.1\%) compared to Fortran2CPP (33.2\%).

Based on this distribution, we focus on the \textit{NoExec} ablation setting, which removes execution-related feedback while preserving the full multi-turn Questioner--Solver interaction. This setting isolates the contribution of execution diagnostics, which provide the richest semantic signals beyond syntactic correctness, while retaining complex debugging trajectories driven by dialogue-level reasoning.

Tables~\ref{tab:codellama_f2c_ablation} and~\ref{tab:codellama_cpp2cuda_ablation} report the performance of CodeLlama-13B-Instruct-hf under this ablation. 
For Fortran2CPP, removing execution feedback leads to only marginal degradation: unit test success decreases from 74.10\% to 73.93\%, compilation success from 88.04\% to 87.88\%, and execution success from 86.50\% to 86.35\%. 
A similar trend is observed for CPP2CUDA, where the \textit{NoExec} setting remains competitive with the full Dialogue configuration across all metrics.

These results suggest that while execution feedback constitutes the most information-rich signal in the dialogue, its removal does not significantly impair performance. This finding aligns with prior observations that state-of-the-art LLMs tend to benefit more from structural and syntactic supervision than from explicit semantic execution feedback~\citep{bitan2025unipar}.

\begin{table*}[t]
\centering
\caption{Ablation study using CodeLlama-13B-Instruct-hf on Fortran2CPP task under different settings.}
\label{tab:codellama_f2c_ablation}
\begin{tabular}{lccc}
\toprule
\textbf{Setting} & \textbf{Unit Test Success} & \textbf{Compilation Success} & \textbf{Execution Success} \\
\midrule
Original      & 276 (42.30\%) & 385 (59.05\%) & 373 (57.21\%)  \\
Code Pair     & 449 (68.90\%) & \textbf{597 (91.56\%)} & 570 (87.42\%) \\
Dialogue      & \textbf{483 (74.10\%)} & 574 (88.04\%) & 564 (86.50\%)  \\
QS-Pair       & 442 (67.80\%) & 592 (90.80\%) & \textbf{576 (88.34\%)}  \\
NoExec     & 482 (73.93\%) & 573 (87.88\%) & 563 (86.35\%) \\

\bottomrule
\end{tabular}
\end{table*}

\begin{table*}[t]
\centering
\caption{Ablation study using CodeLlama-13B-Instruct-hf on CPP2CUDA task under different settings.}
\label{tab:codellama_cpp2cuda_ablation}
\begin{tabular}{lccc}
\toprule
\textbf{Setting} 
& \textbf{Unit Test Success} 
& \textbf{Compilation Success} 
& \textbf{Execution Success} \\
\midrule
Original   & 66  (12.50\%) & 318 (60.23\%) & 243 (46.02\%) \\
Code Pair  & 351 (66.50\%) & 423 (80.11\%) & 414 (78.41\%) \\
Dialogue   & 339 (64.20\%) & 423 (80.11\%) & 404 (76.52\%) \\
QS-Pair 
           & \textbf{363 (68.80\%)} 
           & \textbf{451 (85.42\%)} 
           & \textbf{433 (82.01\%)} \\
NoExec     &  347 (65.72\%) & 426 (80.68\%) & 409 (77.46\%) \\
\bottomrule
\end{tabular}
\end{table*}

\lstset{basicstyle=\footnotesize\ttfamily, breaklines=true, breakatwhitespace=false, columns=fullflexible, tabsize=2, frame=lines, keepspaces=true}
\section{Prompts}
\label{app:prompts-listings}
Below we include the prompt strings used in our pipeline. For different translation tasks, we will attach corresponding examples in prompts.
\subsection*{System Instruction for Questioner: Dual-Agent Translation Pipeline}
\begin{lstlisting}

You orchestrate a two-phase pipeline for code translation and verification.
PHASE A (C++ BENCH FIRST):
1) Ask for a SINGLE-FILE **C++** program that contains:
   - the provided **C++ implementation** (you may refactor into functions),
   - a `main` that constructs deterministic inputs (fixed sizes/seed),
   - **self-checking** that proves correctness,
   - **NO external libs** (NO GoogleTest/Catch2/etc.). OpenMP optional.
2) It must compile with `g++ -fopenmp` (OpenMP optional) and run to completion.
3) It must print exactly one final summary line:
   RESULT_OK checksum=<integer>

PHASE B (TRANSLATE TO CUDA WITH IDENTICAL TEST):
1) After C++ passes locally, ask for a SINGLE-FILE **CUDA** program that:
   - implements the same logic with CUDA kernels,
   - **reproduces the same test scenario** (same inputs, seed, sizes),
   - prints the **EXACT SAME** final summary line:
     RESULT_OK checksum=<integer>
2) If outputs mismatch or any error occurs, you will be given the logs.
   You MUST return a full single-file program in a fenced block that fixes the issue.

\end{lstlisting}

\subsection*{Prompt for Generating Deterministic C++ Benchmark}
\begin{lstlisting}

Please produce a SINGLE-FILE C++ program that both defines the **reference implementation**
AND contains a `main` that builds deterministic inputs and validates outputs.
Requirements:
- No external dependencies or test frameworks.
- If you use OpenMP, allow serial fallback if OpenMP is unavailable.
- End by printing exactly ONE line:
  RESULT_OK checksum=<integer>
Return the entire program in a ```cpp fenced block. Nothing else.

\end{lstlisting}

\subsection*{Prompt for Translating C++ to CUDA under Identical Test Conditions}
\begin{lstlisting}

Translate the validated C++ program into a SINGLE-FILE **CUDA** program with identical logic and the **same test scenario**.
Requirements:
- Keep the same input sizes and data/seed.
- Validate results on host with the same criteria/tolerance.
- Print EXACTLY the same final summary line:
  RESULT_OK checksum=<integer>
Return the entire program in a ```cuda fenced block. Nothing else.

\end{lstlisting}

\subsection*{Concise Question Prompt for Code Translation Queries}
\begin{lstlisting}

I now need to ask you some questions about C++ to CUDA code translation, You need to keep every answer concise.
The first question is: {CPP_Code}

\end{lstlisting}

\subsection*{Prompt for Unit Test Execution and Constraints}
\begin{lstlisting}

{Unit_Test_Request}
I will execute part of the unit test code you gave.
But please note that I cannot download external libraries, so please do not add any external libraries (such as google test) when writing unit testing code.

\end{lstlisting}

\subsection*{Prompt for Removing Comments from Source Code}
\begin{lstlisting}
Help me to delete the comments of the following C++ code:
C++ Code:
{CPP_Code}
C++ Code without comments:

\end{lstlisting}

\subsection*{Prompt for Checking Code Self-Containment}
\begin{lstlisting}

Decide if this C++ snippet is self-contained for immediate test-bench generation.

Self-contained means:  
1. All referenced functions \/ classes are fully defined here **or** are from the standard library.  
2. Adding a minimal `main()` and standard headers lets it compile & link without unresolved symbols.  
3. No external files, network, or special hardware APIs needed.

Return ONLY ``YES" or ``NO".

{CPP_Code}

\end{lstlisting}

\subsection*{Prompt for Initializing Solver with Unit Test Code}
\begin{lstlisting}

C++ Unit test code:
```cpp
{cpp_code}
```

CUDA Unit test code:
```cuda
{cuda_code}
```

\end{lstlisting}

\subsection*{Prompt Template for Repair Intent and Full Program Output}
\begin{lstlisting}
First line = JSON array of 'repair intent tags'; then ONE fenced code block with the FULL single-file program. No other text.
\end{lstlisting}

\subsection*{Prompt for Validating and Aligning C++ and CUDA Test Scenarios}
\begin{lstlisting}

You are given a validated **C++** program and the current **CUDA** program.
First, internally decide whether the CUDA **TEST** (input construction, shapes/sizes, constants, seeds, tolerance,
and the final summary print) is IDENTICAL to the C++ TEST.
- If NOT identical: Modify **ONLY THE TEST SCAFFOLD** of the CUDA program so that it exactly matches the C++ test.
  Do not change kernel math yet. Keep the same final summary line format.
- If identical: Do not change the test; modify **ONLY THE COMPUTATION** (kernels and related glue)
  so that the CUDA output produces exactly this final line: {target_line}.
Strict output format: first line = JSON array of repair intent tags; then ONE ```cuda code block with the full corrected program. No other text.
C++ reference program:
```cpp
{cpp_code}
```
Current CUDA program:
```cuda
{cuda_code}
```

\end{lstlisting}

\subsection*{Prompt Related to CUDA Translation or Validation}
\begin{lstlisting}

The CUDA program's test already matches the C++ test. Do NOT change any test scaffolding (input sizes, seed, I/O, validation).
Modify ONLY the kernels/computation so that the CUDA output produces exactly this final line: {target_line}.
Strict output format: first line = JSON array of repair intent tags; then ONE ```cuda code block with the full corrected program. No other text.
C++ reference program:
```cpp
{cpp_code}
```
Current CUDA program:
```cuda
{cuda_code}
```

\end{lstlisting}

\subsection*{Prompt for Output Equivalence Check Between C++ and CUDA}
\textit{(Code variable: \texttt{ft\_ct\_further\_check})}
\begin{lstlisting}

Answer only `Yes` or `No`: Do the CUDA and C++ programs produce IDENTICAL final summary lines?
C++: {cpp_compile_result}
CUDA: {cuda_compile_result}

\end{lstlisting}

\subsection*{Prompt for Extracting Final Compilable C++ and CUDA Code Pair}
\begin{lstlisting}

Provide the FINAL pair of programs **known to compile and run successfully** and that print identical final summary lines.
Return both in two separate fenced blocks: first ```cpp, then ```cuda. No extra commentary.

\end{lstlisting}

\subsection*{Prompt for Unit Test Modification Based on Execution Output}
\begin{lstlisting}

modify the unit test code based on the outputs and give me the complete modified unit test code to make sure I can compile and run it directly
C++ code result: {cpp_compile_result}
cuda code outputs: {cuda_compile_result}

\end{lstlisting}

\subsection*{Prompt for Delegating C++ to CUDA Translation to Another Agent}
\begin{lstlisting}

Here is my C++ code: {cpp_code}. Now you need to provide a complete question (including code) to the answerer and ask him to translate this C++ code to CUDA code and give you the CUDA code. Don't translate this C++ code by yourself. Ask the answerer to follow the template to start CUDA code with '''cuda and end with '''.

\end{lstlisting}

\subsection*{Prompt for Unit Test Generation or Execution}
\begin{lstlisting}

Here is the answer from the solver: {ser_answer}, you now need to ask the answerer to provide the executable unit-test code for both the original C++ code and the translated CUDA code separately.

Please write the main function for both code and add the unit tests. Add them to C++ code and CUDA code separately.

In the C++ code, you should use 'assert' for the unit-test checking. One example:
assert(your_cpp_function(arg1, arg2) == expected_result);


In the CUDA code, perform the unit-test checking on the host (CPU) after the results have been copied back from the device. Use a pattern similar to:
// Assuming 'h_result' is an array on the host that holds the results
// copied back from the GPU, and 'expected_value' is the expected outcome
// for a specific test case.
if (h_result[i] != expected_value) {{
    printf("Test case failed: assertion failed at index %d!\n", i);
    return 1; // or exit(1)
}}

**Important:**  
* Keep the C++ and CUDA tests completely separate: each must compile and run on its own.
* Provide one same unit test for both C++ and CUDA code.
* The C++ code should start with ```cpp and end with ```.
* The CUDA code should start with ```cuda and end with ```.
* Each program should exit with status 0 (or print a success message) when all tests pass.


\end{lstlisting}

\subsection*{Prompt for Iterative Code Alignment and Unit Test Completion}
\begin{lstlisting}

Help me continue to modify the C++ and cuda codes to ensure that they have the same functions and provide the complete unit test code to make sure I can compile and run it directly (Not only the main code).

\end{lstlisting}

\subsection*{Prompt for Enforcing Yes/No Clarification}
\begin{lstlisting}

Your answer was neither 'yes' nor 'no'. Please provide a clear answer.

\end{lstlisting}

\subsection*{Prompt Related to CUDA Translation or Validation}
\begin{lstlisting}
Help me to translate the following CUDA code to C++ code, don't give any words:
CUDA Code:
__global__ void addKernel(int *c, const int *a, const int *b) {{
    int i = threadIdx.x;
    c[i] = a[i] + b[i];
}}
Translated C++ Code:
void add(int *c, const int *a, const int *b, int size) {{
    for (int i = 0; i < size; ++i) {{
        c[i] = a[i] + b[i];
    }}
}}


Help me to translate the following CUDA code to C++ code, don't give any words:
CUDA Code:
__global__ void saxpy(int n, float a, float *x, float *y) {{
    int i = blockIdx.x * blockDim.x + threadIdx.x;
    if (i < n) {{
        y[i] = a * x[i] + y[i];
    }}
}}
Translated C++ Code:
void saxpy(int n, float a, float *x, float *y) {{
    for (int i = 0; i < n; ++i) {{
        y[i] = a * x[i] + y[i];
    }}
}}


Help me to translate the following CUDA code to C++ code, don't give any words:
CUDA Code:
{CPP_Code}
Translated C++ Code:

\end{lstlisting}

\subsection*{Prompt for CUDA-to-C++ Translation with Explanation (Few-Shot)}
\begin{lstlisting}
Help me to translate the following CUDA code to C++ code by using the following format:
The function of the source CUDA Code is: ...
Translated c++ code is: ...
Explanation: ...

Example:
CUDA Code needs to be translated:
__global__ void addKernel(int *c, const int *a, const int *b) {{
    int i = threadIdx.x;
    c[i] = a[i] + b[i];
}}
Translated C++ Code:
void add(int *c, const int *a, const int *b, int size) {{
    for (int i = 0; i < size; ++i) {{
        c[i] = a[i] + b[i];
    }}
}}



The function of the source CUDA Code is:
The CUDA kernel `addKernel` performs element-wise addition of two integer arrays and stores the result in a third array.  Each GPU thread handles one element, enabling parallel execution.

Translated c++ code is:
#include <cstddef>
void add(int *c, const int *a, const int *b, std::size_t size) {{
    for (std::size_t i = 0; i < size; ++i) {{
        c[i] = a[i] + b[i];
    }}
}}


Explanation:
1. **Parallel vs. serial** - the CUDA kernel runs thousands of threads in parallel; the C++ translation uses a simple `for` loop that executes serially on the CPU.
2. **Thread index** - the GPU index `threadIdx.x` is replaced by the loop variable `i`.
3. **Kernel qualifiers** - `__global__` is removed in C++ because it is specific to GPU kernels; the function becomes an ordinary CPU routine.
4. **Array bounds** - the original kernel relies on the grid configuration to limit `i`; the C++ loop explicitly iterates from `0` to `size-1`.
5. **Headers / types** - C++ version includes `<cstddef>` to get `std::size_t` for portable size handling.

Real Code:
CUDA Code needs to be translated:
{CPP_Code}

\end{lstlisting}

\subsection*{Prompt for CUDA-to-C++ Translation Correctness Verification (Few-Shot)}
\begin{lstlisting}

I will provide you a paragraph of CUDA code and a paragraph of translated C++ code.  
Tell me whether the CUDA code has been **correctly translated** into C++:

* If it **is** correct, answer **"Yes"**.  
* If **not**, answer **"No"** and give a short reason.

CUDA code:
__global__ void addKernel(int *c, const int *a, const int *b) {{
    int i = threadIdx.x;
    c[i] = a[i] + b[i];
}}
Translated C++ Code:
void add(int *c, const int *a, const int *b, int size) {{
    for (int i = 0; i < size; ++i) {{
        c[i] = a[i] + b[i];
    }}
}}

Answer: Yes

CUDA code:
__global__ void saxpy(int n, float a, float *x, float *y) {{
    int i = blockIdx.x * blockDim.x + threadIdx.x;
    if (i < n) {{
        y[i] = a * x[i] + y[i];
    }}
}}
Translated C++ Code:
void saxpy(int n, float a, float *x) {{
    for (int i = 0; i < n; ++i) {{
        x[i] = a * x[i];  // wrong: y is missing and operation differs
    }}
}}

Answer: No
1. Missing parameter **y** in the C++ function.  
2. C++ version overwrites **x** instead of computing **y = a$\cdot$x + y**.

CUDA code:
{CPP_Code}
Translated C++ Code:
{Cpp_Code}
Answer:

\end{lstlisting}

\subsection*{Prompt for Fixing Incorrect C++ Translation Based on Reasons}
\textit{(Code variable: \texttt{Own\_model\_Modify\_code})}
\begin{lstlisting}

The following code translation is not perfect, you need to modify the translated C++ Code based on the reasons.
Original CUDA code:
{CPP_Code}
Translated C++ Code:
{Cpp_Code}
Reasons:
{Reasons}
Modified C++ Code:

\end{lstlisting}

\subsection*{Prompt for Assessing Translated C++ Correctness}
\begin{lstlisting}
Help me to assess if the translated C++ is correct.
                  Source cuda code: {CUDA_code}
                  Translated C++ Code: {Cpp_code}
                  Answer:
\end{lstlisting}

\subsection*{Prompt Related to CUDA Translation or Validation}
\begin{lstlisting}
Help me to modify the translated C++ code based on the reason above, just provide the modified C++ code, don't give any words:
C++ Code:
#include <stdio.h>
int a[100][100];
int main(){{
    int i, j;
    #pragma omp parallel for collapse(2)
    for (i = 0; i < 100; ++i){{
        for (j = 0; j < 100; ++j){{
            a[i][j] = a[i][j] + 1;
        }}
    }}
    return 0;
}}


Help me to modify the translated C++ code based on the reason above, just provide the modified C++ code, don't give any words:
C++ Code:
#include <stdio.h>
#if (_OPENMP < 201511)
#error "An OpenMP 4.5 compiler is needed to compile this test."
#endif

int a[100][100];
int main(){{
    int i, j;
    #pragma omp parallel
    {{
        #pragma omp single
        {{
            #pragma omp taskloop collapse(2)
            for (i = 0; i < 100; ++i){{
                for (j = 0; j < 100; ++j){{
                    a[i][j] += 1;
                }}
            }}
        }}
    }}
    printf("a[50][50] = %d\n", a[50][50]);
    return 0;
}}


Help me to modify the translated C++ code based on the reason above, just provide the modified C++ code, don't give any words:
C++ Code:

\end{lstlisting}

\subsection*{Prompt for Fixing C++ Compilation Errors with One-Shot Example}
\begin{lstlisting}
I am trying to translate a paragraph of CUDA code to C++ code, but the translated C++ code cannot pass the compiler.  
Please modify the C++ code so that it compiles successfully.  
**Return only the modified C++ code with no explanations.**

Source CUDA Code:
__global__ void addKernel(int *c, const int *a, const int *b) {{
    int i = threadIdx.x;
    c[i] = a[i] + b[i];
}}

Translated C++ Code:
void add(int *c, const int *a, const int *b) {{
    // BUG: missing loop and index declaration
    c[i] = a[i] + b[i];
}}

Modified C++ Code:
void add(int *c, const int *a, const int *b, int size) {{
    for (int i = 0; i < size; ++i) {{
        c[i] = a[i] + b[i];
    }}
}}


Source CUDA Code:
{CPP_Code}
Translated C++ Code:
{Cpp_Code}
Modified C++ Code:

\end{lstlisting}

\subsection*{Prompt for Fixing C++ Compilation Errors (Zero-Shot)}
\begin{lstlisting}
I am trying to translate a paragraph of CUDA code to C++ code, but the translated C++ code cannot pass the compiler.  
Please modify the C++ so it **compiles successfully**.  
Return **only** the modified C++ code with no explanations.

Source CUDA Code:
__global__ void addKernel(int *c, const int *a, const int *b) {{
    int i = threadIdx.x;
    c[i] = a[i] + b[i];
}}

Translated C++ Code:
void add(int *c, const int *a, const int *b) {{
    // BUG: missing loop and index
    c[i] = a[i] + b[i];
}}

Source CUDA Code:
{CPP_Code}
Translated C++ Code:
{Cpp_Code}

\end{lstlisting}

\subsection*{Prompt for Iterative C++ Compilation Error Repair}
\begin{lstlisting}

The compiler is throwing errors. The error report is: {reason}. Please help me to continue modifying the C++ code. Just write out the modified C++ code based on the error report, DO NOT write other words! New C++ Code:

\end{lstlisting}

\subsection*{Prompt for Unit Test Generation or Execution}
\begin{lstlisting}

I want you to help me choose suitable code for a unit test. I will provide
two snippets: one in CUDA and one in C++.

Tasks
-----
1. Check whether the two snippets have the same value based input and output parameters.
2. If they do not, reply "False".
3. If they do, output a Google Test skeleton of the form

TEST(MyLib, MyKernel_test) {{
    // ---- prepare identical inputs ----
    /* ... */

    // ---------- C++ reference ----------
    cpp_function_name(/* host args */);

    // ---------- CUDA kernel launch ----------
    /* allocate device memory, copy inputs, launch, copy outputs back */

    EXPECT_EQ(/* C++ result */, /* CUDA result */);
}}

Notes
-----
* Wrap literal braces in your output as double braces {{ }} so they are not interpreted as format placeholders.
* When calling the CUDA kernel, wrap it in a simple host function or launch it directly with <<<grid, block>>>.
* Use EXPECT_EQ, EXPECT_FLOAT_EQ, or EXPECT_NEAR as appropriate.

Examples
========

Example 1 -> False
------------------
CUDA code:
__global__ void badIndex(float* a) {{
    int idx = 1.5f;  // illegal float index
    a[idx] = 0.0f;
}}

C++ code:
int main() {{
    float a[10];
    a[1] = 0.0f;
    return 0;
}}

Answer:
False

Example 2 -> Unit-test skeleton
-------------------------------
CUDA code:
__global__ void saxpy(int n, float a,
                      const float* x, float* y) {{
    int i = blockIdx.x * blockDim.x + threadIdx.x;
    if (i < n) {{
        y[i] = a * x[i] + y[i];
    }}
}}

C++ code:
void saxpy_cpu(int n, float a,
               const float* x, float* y) {{
    for (int i = 0; i < n; ++i) {{
        y[i] = a * x[i] + y[i];
    }}
}}

Answer:
TEST(SaxpyLib, Saxpy_test) {{
    const int N = 1024;
    const float A = 2.0f;
    float hx[N], hy_cpp[N], hy_cuda[N];

    for (int i = 0; i < N; ++i) {{
        hx[i]      = static_cast<float>(i);
        hy_cpp[i]  = static_cast<float>(i * 0.5f);
        hy_cuda[i] = hy_cpp[i];
    }}

    saxpy_cpu(N, A, hx, hy_cpp);  // C++ reference

    float* dx;
    float* dy;
    cudaMalloc(&dx, N * sizeof(float));
    cudaMalloc(&dy, N * sizeof(float));
    cudaMemcpy(dx, hx,      N * sizeof(float), cudaMemcpyHostToDevice);
    cudaMemcpy(dy, hy_cuda, N * sizeof(float), cudaMemcpyHostToDevice);

    dim3 block(256);
    dim3 grid((N + block.x - 1) / block.x);
    saxpy<<<grid, block>>>(N, A, dx, dy);
    cudaMemcpy(hy_cuda, dy, N * sizeof(float), cudaMemcpyDeviceToHost);

    for (int i = 0; i < N; ++i) {{
        EXPECT_FLOAT_EQ(hy_cpp[i], hy_cuda[i]);
    }}

    cudaFree(dx);
    cudaFree(dy);
}}

Example 3 -> no code provided
-----------------------------
CUDA code:

C++ code:

Answer:


Real code
---------
CUDA code:
{cpp_code}

C++ code:
{cpp_code}

Answer:

\end{lstlisting}

\subsection*{Prompt for Code Modification with Debugging and Dependency Instructions}
\begin{lstlisting}
Please modify the code and give the modified complete code, make sure all the functions are within a file and I will re-run the code.
1. You can add debugging statements if needed.
2. If there is a need for external library installations, please let me know the appropriate pip command by enclosing them in ```sh ```
\end{lstlisting}

\subsection*{Prompt for Judging Test Correctness from Execution Output}
\begin{lstlisting}
Please judge whether the test code you just gave is correct based on the output of the code execution. Just Answer: 'Yes' or 'No'. 
\end{lstlisting}

\subsection*{Prompt for Extracting Corrected Function Code Without Tests}
\begin{lstlisting}

Give me the correct modified function code (without the test code) based on your last unit test code.

\end{lstlisting}

\subsection*{Prompt for General Code Fix and Re-Execution}
\begin{lstlisting}

Please go ahead and modify the code to make sure it can run correctly.
You should make sure all the functions are within a file and I will re-run the code.

\end{lstlisting}

\subsection*{Prompt for Self-Verification with Mock Data}
\begin{lstlisting}

Could you help verify whether your code can run correctly?
1. If needed, You could create some mock data or files to assist with this. But note that whether you create new data or create a new file and write the data to it, these operations need to be done in same python file.
2. I will help you to install the related packages, you just need to tell me how install the package you need by using ```sh ... ```.
3. Our goal is to verify that the function works correctly. So you need to make sure you provide me with a complete python code rather than providing some simplified version of it.

\end{lstlisting}

\subsection*{Prompt Paraphrases for Requesting Code Execution Verification}
\begin{lstlisting}

To confirm the code functions properly, we should execute it and check its performance.
Let's test the code to make sure it operates as expected.
To verify that the code is functioning correctly, let's run a test.
Let's execute the code to validate its proper functioning.
To make certain our code is running right, we should perform a test.
Let's initiate a run to confirm that the code works as intended.
To ascertain the code's effectiveness, we must test it.
We need to run the code to ensure it meets our standards.
Let's check the code's functionality by running it.
To guarantee the code's accuracy, testing it is essential.
We should execute the code to verify its accuracy.
Let's run the code to make sure everything is functioning properly.
To ensure flawless operation, we need to test the code.
Let's operate the code to check its effectiveness.
We should validate the code's performance by running it.
Let's put the code through a run-test to ensure it works correctly.
To be certain of the code's operation, we need to run it.
Running the code will help us verify its proper function.
Let's test run the code to check for any issues.
To confirm code reliability, let's execute it now.
We should run a trial to test the code's functionality.
Let's activate the code to ensure it's working as it should.
Running the code will confirm its efficiency and correctness.
We need to execute the code to confirm that it performs correctly.
To make sure the code is error-free, let's run a verification test.
Let's run the code to determine if it functions correctly.
We should check the code by running it to ensure its efficacy.
Let's perform a test run to validate the code's functionality.
To ascertain code performance, executing it is necessary.
We must run the code to ensure it operates efficiently.
To verify the code's success, let's give it a run.
Running the code will allow us to confirm its functionality.
We need to execute the code to see if it's working properly.
Let's initiate a test run to ensure the code is effective.
To confirm the code's operational success, testing it is crucial.
Let's deploy the code to check its working condition.
Running the code is essential to verify its proper execution.
We should operate the code to confirm its capabilities.
Let's activate a test run to ensure the code functions well.
To make sure the code performs its intended functions, we need to run it.
Running the code will help us ensure it meets functional requirements.
We should test the code to confirm that it executes properly.
Let's perform a functional test to ensure the code is running correctly.
To check the code's precision, let's run it.
We must initiate a test to verify the code's functionality.
To determine if the code is error-free, we should run it now.
Running the code is the best way to ensure its accuracy.
Let's conduct a run to verify that the code operates as it should.
We need to run the code to check its functionality and reliability.
Let's execute the code to test its overall performance.

\end{lstlisting}

\subsection*{Prompt Paraphrases for Asking the Model to Self-Test Code}
\begin{lstlisting}

Can you check if your code runs as expected?
Could you test your code to see if it functions properly?
Would you mind confirming that your code operates correctly?
Can you please verify the functionality of your code?
Could you ensure that your code is executing correctly?
Would you verify whether your code is functioning properly?
Can you determine if your code performs well?
Could you assist in checking if your code runs smoothly?
Can you help confirm your code's correctness?
Would you be able to test if your code is working right?
Can you examine your code to ensure it operates as intended?
Could you validate that your code works properly?
Would you mind running a test on your code to verify its performance?
Can you see if there are any issues with how your code runs?
Could you take a look to confirm your code functions correctly?
Would you mind ensuring your code's accuracy?
Can you help determine if your code is error-free?
Could you check for any flaws in your code's operation?
Would you be able to confirm the operational functionality of your code?
Can you make sure that your code is free of errors?
Could you perform a quick check to see if your code runs correctly?
Would you test your code to ensure it's executing properly?
Can you verify that your code meets the expected standards?
Could you assist by verifying your code's performance?
Would you mind checking if your code executes without issues?
Can you confirm the reliability of your code?
Could you please ensure that your code works as it should?
Would you assess whether your code runs effectively?
Can you help verify your code's operational correctness?
Could you double-check the functioning of your code?
Would you mind verifying that your code operates without problems?
Can you look into whether your code functions as planned?
Could you help confirm that your code performs as needed?
Would you be willing to check your code for proper operation?
Can you ensure that your code executes as expected?
Could you run a diagnostic to see if your code is working correctly?
Would you mind testing your code for functionality?
Can you verify the accuracy of your code's execution?
Could you provide assurance that your code is running properly?
Would you verify if your code is up to performance standards?
Can you check your code for any operational errors?
Could you run a trial on your code to confirm it functions correctly?
Would you conduct a check to ensure your code is accurate?
Can you help make sure your code is functioning correctly?
Could you assess if your code is performing as expected?
Would you mind giving your code a run-through to check its correctness?
Can you verify if your code is functioning up to standards?
Could you please test your code for correct operation?
Would you be willing to help ensure that your code is running smoothly?
Can you confirm if your code meets all functional requirements?
Make sure the code runs as intended.
Ensure that the code functions properly.
Verify the correct operation of the code.
Confirm that the code performs as expected.
Double-check the execution of the code.
Test the code to see that it works correctly.
Ascertain that the code behaves as it should.
Check that the code operates correctly.
Ensure the code executes without errors.
Validate the functionality of the code.
Review the code's performance to ensure accuracy.
Monitor the execution to confirm it's functioning properly.
Examine the code to make sure it runs smoothly.
Assess whether the code meets the expected outcomes.
Certify that the code is executing as planned.
Scrutinize the code for proper execution.
Check off that the code works as designed.
Inspect the code for correct operation.
Reconfirm that the code operates as intended.
Authenticate the correct running of the code.
Proof the code to ensure it's working correctly.
Cross-verify the execution of the code.
Make certain the code is functioning correctly.
Confirm the proper execution of your code.
Ensure your code executes as planned.
Verify that your code runs correctly.
Double-check that your code is functioning properly.
Make sure your code performs as expected.
Check that your code operates smoothly.
Validate that your code meets the operational criteria.
Ensure the code does what it's supposed to do.
Test to ensure the code's functionality.
Confirm that the code is error-free upon execution.
Ensure that the code delivers the expected results.
Double-check the code's results for accuracy.
Confirm that the code's output is as anticipated.
Make sure the code completes without issues.
Ensure that the code's performance is up to standard.
Confirm the reliability of the code during execution.
Make sure the code's logic performs correctly.
Verify the outcome of the code's execution.
Check for any discrepancies in the code's operation.
Ensure the code executes as it is supposed to.
Confirm the stability of the code upon execution.
Test the code thoroughly before finalizing.
Make certain the code executes without any hiccups.
Validate the precision of the code's operation.
Check that the code complies with the requirements.
Recheck the code's functionality for assurance.
Ensure that the code achieves the intended functionality.
Monitor the code to ensure it executes flawlessly.
Evaluate the code's execution to confirm correctness.
Ascertain that the code meets performance standards.
Review the code to ensure it fulfills its purpose.
Confirm that the code executes according to the plan.
Make sure the code is free from execution errors.
Ensure that the code's execution aligns with expectations.
Cross-check to confirm the code's proper functionality.
Validate that the code operates as it should.
Confirm that the code's process is correct.
Make certain the code's results are accurate.
Ensure that the code runs efficiently.
Verify that the code's performance is satisfactory.
Check that the code is functioning as it should.
Test the code's execution for any potential issues.
Make sure the code functions as intended under all conditions.
Confirm the integrity of the code's operations.
Ensure that the code performs effectively.
Review the code execution for any anomalies.
Validate the effectiveness of the code's execution.
Double-check the code for flawless performance.
Ensure that the code meets all operational expectations.
Confirm that the code handles all scenarios correctly.
Verify the code's effectiveness in real conditions.
Check the code's consistency in execution.
Make certain that the code adheres to the expected behavior.
Ensure the code's compliance with the specifications.
Confirm the code's capability to perform as necessary.
Monitor the code's performance for stability.
Ascertain that the code is ready for deployment.
Make sure the code satisfies all functional requirements.
Ensure that the code executes without deviations.
Confirm the code's readiness for operational use.
Test the code for dependability during execution.
Verify the code's robustness in various environments.
Check that the code performs efficiently.
Confirm that the code is up to professional standards.
Ensure the code's operations are in full effect.
Reaffirm the code's readiness for live environments.
Make sure the code is optimized for performance.
Confirm the code's suitability for the intended tasks.
Double-check the code for operational accuracy.
Ensure the code meets the quality standards.
Verify the code's readiness for full-scale use.
Confirm that the code maintains consistency.
Make certain the code delivers on its promises.
Ensure the code's functionality before release.
Confirm that the code functions optimally.
Double-check to ensure the code's successful execution.
Make sure the code's behavior matches the documentation.

\end{lstlisting}

\subsection*{Prompt Paraphrases for Reporting Successful Test Passage}
\begin{lstlisting}

Our code has passed all the tests successfully, here's the code:
The code has successfully cleared all the tests, here it is:
We've successfully passed all tests with our code, here's the code:
All tests have been successfully passed by our code, here is what we wrote:
Our program succeeded in all the tests, here's the code:
Our code has cleared all its tests successfully, here's the code:
We've completed all tests successfully, here's our code:
Our coding tests were all passed successfully, here's our code:
Every test has been successfully passed by our code, see it here:
Our code achieved success in all the tests, here it is:
Successfully, our code has passed all tests, here's what we developed:
Our code met all the test criteria successfully, here is the code:
Our code excelled in all the tests, here's the code:
All the tests have been cleared by our code, here it is:
Our software has successfully passed all the testing, here's the code:
Every test was a success for our code, here's the code:
We've passed all the required tests with our code, here it is:
Our code was successful in all the tests, here is our work:
The tests were all successfully passed by our code, here it is:
Our code sailed through all the tests, here's the code:
Our code has surpassed all testing successfully, here it is:
We've successfully navigated all tests with our code, here's the result:
All required tests were successfully passed by our code, see it here:
Our code has triumphed in all the tests, here's our code:
Every test has been successfully conquered by our code, here it is:
Our application passed all tests successfully, here is the code:
The code has proven successful in all tests, here it is:
Our code has come through all the tests with success, here's the code:
Our programming successfully passed all examinations, here's our code:
We've mastered all the tests with our code, here it is:
Our system has successfully passed all tests, here is the code:
The code has been successful in all tests, here is our work:
Every testing hurdle was successfully cleared by our code, here it is:
We've cleared all tests with our code successfully, here's the code:
Our code has been vetted and succeeded in all tests, here it is:
All tests have been met with success by our code, here's the code:
Our project successfully passed all the test phases, here is our code:
Successfully, our code has conquered all the tests, here's the output:
Our code managed to pass all tests successfully, here's the code:
The code has fulfilled all test conditions successfully, here it is:
Our code was tested and passed all assessments successfully, here's the code:
Every test was smoothly passed by our code, here's what we coded:
Our code nailed all the tests successfully, here's our work:
We have successfully completed all tests with our code, here it is:
Every test criteria was met successfully by our code, here's the code:
Our script passed all tests without fail, here is the code:
Our code stood up to all tests and passed, here's the code:
All testing barriers were successfully broken by our code, here it is:
Our code has flawlessly passed all the tests, here's our script:
Our code has been validated through all tests successfully, here's the code:

\end{lstlisting}

\end{document}